\documentclass[ %
longbibliography,reprint,amsmath,amssymb,aps,floatfix,
]{revtex4-1}

\usepackage{graphicx}% Include figure files
\usepackage{dcolumn}% Align table columns on decimal point
\usepackage{bm}% bold math
\usepackage[normalem]{ulem}
\usepackage{amsmath}
\usepackage{gensymb}

\newcommand{\be}{\begin{equation}}
\newcommand{\ee}{\end{equation}}
\newcommand{\bea}{\begin{eqnarray}}
\newcommand{\eea}{\end{eqnarray}}
\newcommand{\bean}{\begin{eqnarray*}}
\newcommand{\eean}{\end{eqnarray*}}

\begin{document}

\preprint{APS/123-QED}

\title{A universal rank-order transform to extract signals from noisy data}% Force line breaks with \\

\author{Glenn Ierley}
\email{grierley@ucsd.edu}
 \affiliation{Mathematics Department, Michigan Technological University\\
 1400 Townsend Drive, Houghton MI 49931}
 \affiliation{Scripps Institution of Oceanography, UC San Diego (emeritus)}
 
\author{Alex Kostinski}%
\email{kostinsk@mtu.edu}
\affiliation{
 Physics Department, Michigan Technological University\\
 1400 Townsend Drive, Houghton MI 49931 
}

\date{\today}

\begin{abstract}
We introduce an ordinate method for noisy data analysis, based solely on rank information and thus insensitive to outliers. The method is nonparametric, objective, and the required data processing is parsimonious. Main ingredients are a rank-order data matrix and its transform to a stable form, which provide linear trends in excellent agreement with least squares regression, despite the loss of magnitude information. A group symmetry orthogonal decomposition of the 2D rank-order transform for iid (white) noise is further ordered by principal component analysis. This two-step procedure provides a noise ``etalon'' used to characterize arbitrary stationary stochastic processes. The method readily distinguishes both the Ornstein-Uhlenbeck process and chaos generated by the logistic map from white noise. Ranking within randomness differs fundamentally from that in deterministic chaos and signals, thus forming the basis for signal detection. To further illustrate the breadth of applications, we apply this ordinate method to the canonical nonlinear parameter estimation problem of two-species radioactive decay, outperforming special-purpose least square software. It is demonstrated that the method excels when extracting trends in heavy-tailed noise and, unlike the Thiele-Sen estimator, is not limited to linear regression. Lastly, a simple expression is given that yields a close approximation for signal extraction of an underlying generally nonlinear signal. 
\end{abstract}

\pacs{Valid PACS appear here}% PACS, the Physics and Astronomy
\maketitle

\section{Preview and Introduction}

We report on a discovery of a rank-based method that appears remarkably versatile and robust with respect to the nature of noise. This is so because the method is ordinal, nonparametric, and therefore distribution-independent. Throughout the paper, the performance of the method is compared to leading nonparametric tests and software, using real as well as synthetic data, where exact results are known.  As new results abound, but the most important ones appear in later sections (\ref{sec:svd} and on), we begin with the slightly unconventional device of an annotated table of contents to orient the reader. 

In Section \ref{sec:deriveQ} we introduce and motivate the initial construction of our method (dubbed there the $Q$ transform) in a simple setting: we begin by solving for the long term warming trend buried in a fluctuating time series of daily low temperature. The same quantity later identified as a diagnostic for signal detection is simultaneously here used for signal extraction by means of parameter estimation (here, the slope). Agreement with the least squares method is excellent. This is quite surprising, given that the method retains no magnitude information whatsoever, only rank. This is a setting with few outliers, where the two approaches generally agree.

In Section \ref{sec:Qapprox} we propose a continuous approximation for $Q$, in terms of which one can understand $Q$ as a simple 2-D integral transform. This formulation facilitates accurate approximation of various basic results (Figs. 2, 3, 7, and \ref{fig:qzero}) with algebraic forms that are more transparent in meaning than the equivalent discrete forms. 

In Section \ref{sec:stats} we introduce two statistical metrics used for confidence tests, characterize their distributions, and give an asymptotic approximation for the scaling of each. The case of correlated noise is also considered.  

In Section \ref{sec:svd} we give a universal representation of the $Q$ transform for all distributions of iid (white) noise. Key is a five term exact orthogonal decomposition based on planar group character, applied to all realizations of $Q$ in an ensemble. Principal component analysis (PCA) is used on each of the resulting group ensembles. The lifting of the original 1D time series to the 2D rank-order space of $Q$  -- ``order'' here taken as time-like, but generally representing any serial independent variable -- establishes a link between $Q$ modes and corresponding ordered patterns of (sample) nonstationarity in mean and variance. As a consequence, $Q$-based slope estimates from Section \ref{sec:deriveQ} for long term trends are unaffected by trends in variance. These ideas are further developed in Section \ref{subsec:fingerprint}, where a new metric is developed for characterizing stochastic processes, offering a prejudice-free means of selecting a model for experimental data. 

In Section \ref{sec:logistic} we address a detection problem where the signal is a chaotic series generated by the logistic map. Our method, which makes no assumptions about the functional form of the underlying signal, readily distinguishes the presence of chaotic signals, whether alone or in combination with white noise. 

In Section \ref{sec:regress} we consider the canonical nonlinear parameter estimation problem for noisy two-species radioactive decay \cite[Chapter 8]{Bevington92}.  In this problem of quantitative signal extraction our method outperforms special-purpose least square software by stably retrieving both decay rates. 

In Section \ref{sec:dqdt} we introduce a heuristic approximation for extracting a complex signal up to within a linear re-scaling by simple differentiation of the transformed field.

In Section \ref{sec:heavytail}  two data sets with distributions of infinite mean and variance noise are explored. 
For such distributions,  the Theil-Sen nonparametric method is commonly used, but is limited to linear regression. Our transform also succeeds for the linear problem but extends to arbitrary functional forms and multilinear settings as well.

In Section \ref{sec:extension} we close with an extension of the method to unequally spaced time series. We develop the theoretical basis for error analysis and apply it to linear regression, hence accounting for the otherwise enigmatic agreement of the linear fits exhibited in Section II.

\subsection*{Signal Detection}

To place the $Q$ transform within the existing literature on time-series analysis, consider signal detection first where statistical signal processing is, perhaps, the natural setting. Here one devises a test statistic and selects an operating threshold \cite{kay1998}. Performance as judged by false positives/negatives is typically characterized with, e.g., a 
receiver operating characteristic curve. If used in the time domain, most such detectors are local; they use a single realization consisting of short segment of the signal to evaluate the test statistic and assign a score. The resulting sequence of statistics for the entire time series identifies discrete intervals where signal is likely present. In \cite{bandt2002} information entropy was proposed as such a test statistic. Initially applied to detection for speech and deterministic chaos, it has been widely used, e.g., \cite{garland2018,fischer2017}. 

Like the $Q$ transform, this information entropy is also rank-based and nonparametric. However, it is a ``local'' measure whereas $Q$ is ``global''. By global we mean that evaluation of $Q$ relies on a significant number of trials to accumulate sufficient statistics about the parent noise distribution. This global approach performs well for detection at a poor signal-to-noise-ratio (SNR) when local methods would fail. The weakness is that $Q$ cannot be obtained from a single trial. One can liken the {\em global} character of the approach to a spectrogram-based signal detection in the frequency domain (see e.g.\ \cite{helble2012}) but, where the latter are usually energy detectors (or other power-law), $Q$ detection is based solely upon rank.

\subsection*{Signal Extraction}

Turning to the subsequent problem of signal extraction,  this is often accomplished by some variant of a least squares minimization, and a vast literature supports this approach. For example, when errors are identically and independently distributed (iid) Gaussian random variables, ordinary least squares  is the maximum likelihood estimator, e.g., see \cite{Lupton1993,Bevington92}. However, nonstationary variance is ubiquitous in data analysis and so is lack of independence. These complications could be addressed with generalized least squares using a weight matrix equal to the inverse of the covariance matrix, $\Omega$, when the covariance of the fluctuations is known. In practice $\Omega$ must be estimated. For this ``feasible generalized least squares'' it is difficult to assess the effect of error with empirical weights. Correlated non-stationary noise is often heavy-tailed, e.g.\ see \cite{bardou2002levy} for numerous examples in atomic physics, and outliers are then a serious problem for least squares. Rank-based methods need no empirical weights for such complications.  Two species radioactive decay is a case where the least square error itself -- nonlinear in the parameters -- may fail as a penalty function, while our rank-based measure proves robust. 

For parameter estimation, one chooses a representation for the solution, either specific to the application, as with exponential decay, or a generic form such as a polynomial expansion. The coefficients in the functional form are determined by a minimization procedure. 

For non-parametric signal extraction, we make no assumption about form apart from spectral separation. The natural comparison for a deterministic signal buried in noise is a moving average convolution, with the stencil of weights ranging from a simple boxcar to a precisely designed filter for impulse response.  Such filters are applied to single realizations whereas $Q$ needs an ensemble.

In summary: in the realms of both detection and extraction, to the best of our knowledge there are no methods that are rank-based, nonparametric, and global in the sense defined above. 

\section{Description of the $Q$ transform and trend extraction \label{sec:deriveQ}} 

We chose climate as a setting to initially motivate and illustrate the method, but several other contexts will be provided throughout the paper. Nonparametric statistics have been used in climate physics, e.g., record-breaking statistics have been employed to infer a variety of trends from temperature time-series \cite{Benestad2003, Benestad2004, Meehl2009, Anderson.Kostinski2010}. Such nonparametric and distribution-free methods are, indeed, an alternative to the various least squares methods.  However, to the best of our knowledge, up to now only record lows and record highs have been used in the climate context, e.g., \cite{coumou2012decade}.  Here, we are guided by the simple thought that the entire rank information and not just its first and last element, ought to be used in nonparametric analyses and our results buttress this claim. Throughout this paper, the $i$th entry in a time-series, $x_{i}$, is assigned rank $r$ if it is the $r^{th}$ lowest value of the entire sequence when sorted by magnitude.  For example, the high or low daily temperature at a particular location, $T = T(t)$ is sorted and ranked below, and we track the year of origin (order, $t$). Hence, the ``rank-order'' in the title.  We note in passing that rank is not always uniquely defined as ties occur. The subject of ties merits a paper of its own paralleling considerations raised e.g.\ in \cite{edery2013record}. To circumvent this problem we either assign fractional rank or add white noise. In this paper we examine data sets of daily high temperatures from the Global Historical Climatology Network (GHCN) and raw monthly mean temperatures from the Berkeley Earth repository. 

\begin{figure*}[ht]
\centerline{\includegraphics[height=4.45in]{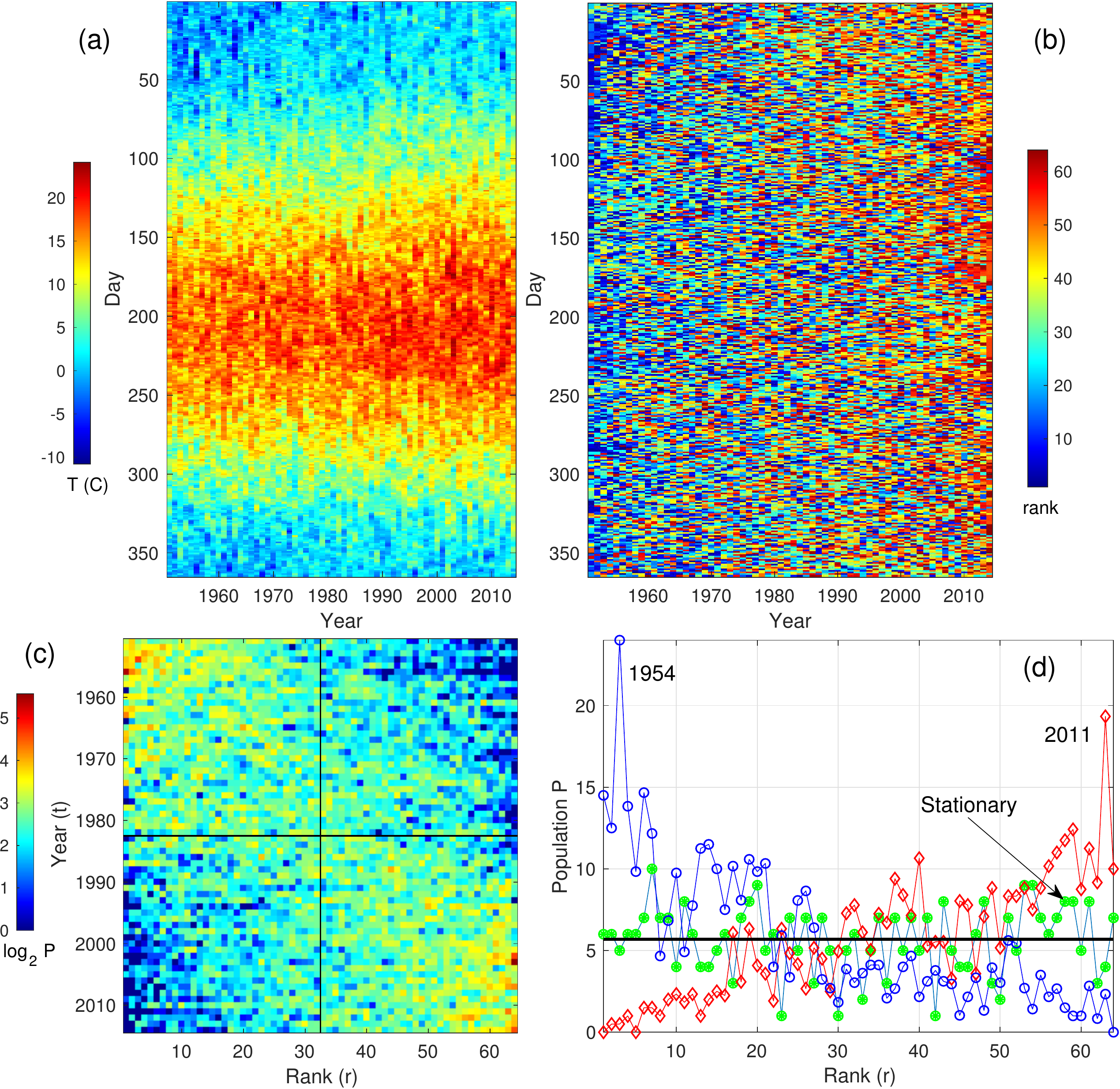}} 
  \caption{{\bf Data processing, illustrated on weather station SZ000009480 (GHCN) Lugano, Switzerland:}  (a) Daily high temperature values displayed as a day (row) and year (column) matrix; (b) Same data but with only the daily rank recorded in rows; (c) the $64 \times 64$ rank-year square matrix $P$ where each entry is the color-coded number of occurrences of that particular rank that year (``occupation number''); the combined population of upper left and lower right quadrants (defined by the cross-hairs) is 14083 whereas the combined population of the lower left and upper right is 9277, with the expected population for stationary climate (365 x 64)/2 = 11680.  This quadrupolar asymmetry constitutes a warming signal; (d) 1954 and 2011 population vs.\ rank: the near reflection symmetry between the red (diamond) and blue (circle) curves is evident and compatible with warming, with 75\% of the 1954 population in the bottom half of ranks and 75\% in the top half for 2011. The 1954 and 2011 maxima are at ranks 3 and 63, respectively. Considering the mean of $\mu = 365/64 = 5.70$ and the Poisson pdf (valid for iid climate, see text), 24 occurrences of rank 3 in the year 1954 are exceedingly unlikely for a stationary climate ($7 \times 10^{-7})$. On the other hand, the number of record highs in 2011 is 10 which is plausible ($3\%$) for a stationary climate (occurring once also in the green (asterisk) curve, which is one realization of a stationary climate). Hence, {\em most of the essential information here is contained in the intermediate ranks}. The argument is stronger yet for autocorrelated data.}\label{fig:1abcd}
\end{figure*}

As a specific example, consider the GHCN weather station SZ000009480 (Lugano, Switzerland).  Color is used in Fig. \ref{fig:1abcd}(a) to display daily high temperature values as a day (row) and year (column) matrix.  The seasonal variability is apparent, e.g., almost everything is red around day 180 (summer).  In Fig. \ref{fig:1abcd}(b) we display the same data but with daily rank recorded in rows: all magnitude information has been discarded and all data are now integer-valued.  The appearance is fine-grained, reminiscent of ``salt-and-pepper'' noise. The central finding of this paper is that the trend information content of \ref{fig:1abcd}(a) and \ref{fig:1abcd}(b) is almost identical (for a large data set), despite the total loss of magnitude data.  This is appealing, as ranking is affected neither by outliers, nor any monotonic transformation of temperature data, e.g., a logarithm \cite{ThompsonMacdonald1991}, nor by occasional gaps in data as shown below. To introduce the approach, we begin by re-packaging the day/year rank matrix data.

Disregarding the  dependence (for the moment), let us view the daily temperature values in Fig. 1(b) as independent random trials, indexed by year. For example, among the 365 trials during year 1951, nine record low (rank 1) values occurred, that is, lower than any of the 63 subsequent values (1952-2014) for that day. Given the independent trials perspective, the essential information can be distilled to just three numbers: only the year, the rank, and the ``population'' of that rank need be preserved. The order of occurrence of the nine ``events'' is superfluous as the events are indistinguishable (because the trials are independent and, for the moment, seasonality is not a concern). Therefore, the input data matrix of Fig. 1(b) can be condensed. Guided by this observation, we let the rank be an independent variable and construct a 64 x 64 rank-order square matrix $P$ as shown in Fig. 1(c) where each entry is the ``occupation number'' or the number of occurrences for that particular rank and year.  The total population of the $P$-matrix is $365 \times 64 $. $P$ is integer-valued, invariant with respect to temperature offset, and the total population of each row and column is $365$. More generally for $P$ the range is [$0,n_t$] where $n_t$ is the number of trials (here days). 

Note that the entries of $P$ are not evenly distributed among the quadrants defined by the cross-hairs in Fig. 1(c). Whereas the combined population of upper left and lower right quadrants is 14083, that of lower right and upper left is 9277. The expected population, given a stationary climate, is $(365 \times 64)/2 = 11680$. This nonstationarity of $\approx 20.6 \%$ is of overwhelming statistical significance, and we use this message in the data to work towards an objective, assumption-free definition of a warming signal. The extreme case of a pure warming trend with no variability results in a $P$ which is a multiple of the identity matrix, with a pre-factor $n_t$. By contrast, consider an ensemble of stationary climate realizations.  For a given time series of 64 years, any entry is equally like to be the hottest (record-breaking) and shuffling these entries does not change the statistics, because of independence \cite{foster1954distribution}. Then, in the limit, ensemble-averaged populations of all ranks of a given row of $P$ (fixed time) should be equal and the matrix $P$ should approach perfect uniformity (all matrix elements equal, $P = const$).

\begin{figure*}[ht]
 \centerline{\includegraphics[width=5in]{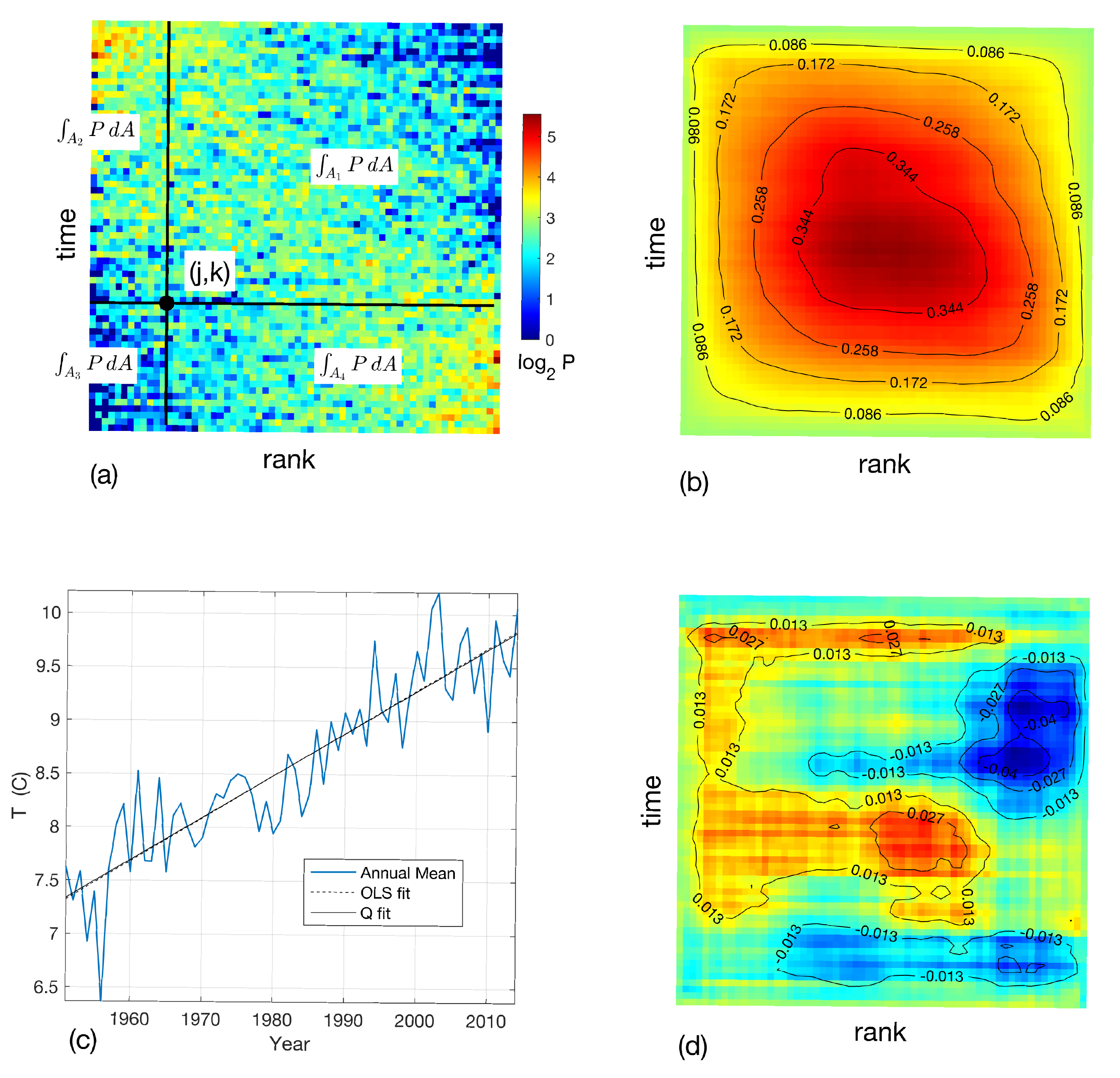}}
 \caption{{\bf $P$ to $Q$ transformation and resulting trend, for SZ000009480 (GHCN), Lugano, Switzerland:}  (a) cross-hairs centered at $(j,k)$ grid element; partitioning of $P$, used to compute the $(j,k)$ element of $Q$; (b) $Q$ computed via (\ref{eq:qdiscrete}), reveals a prominent warming pattern;  (c) the linear trend, obtained by annulling the matrix element average, $\langle Q \rangle$, ($2.4875 \degree$C) is nearly identical to the standard LS fit ($2.5165 \degree$C); (d) shows the residual $Q$ computed from the raw temperature record after linear detrending (color scale expanded from that for (b) to preserve detail). Note the large scale residual pattern, which is an order of magnitude smaller than the original $Q$. All plots of $Q$ throughout the paper employ dark blue and dark red to denote bounds of $[- \max | Q | , \max |Q| ]$ respectively. Henceforth pale green will thus indicate zero in these plots.}\label{fig:2abc}
\end{figure*}

To gain further insight into the meaning of the $P$-signal consider an early and late year, namely, 1954 (order 4) and 2011 (order 61), displayed as a histogram versus rank in Fig. \ref{fig:1abcd}(d). Observe that, for a steady climate, rank occupation numbers, approximated as independent trials (akin to classical particles), obey Poisson statistics: $p(n) = (\mu^{n}/n!) e^{-\mu}$, with $\mu = 365/64 = 5.70$ being the average population per rank and $\sigma = (365/64)^{1/2} = 2.39$ the standard deviation. Hence, we expect $\approx 6 \pm 2$ as the green curve (labeled stationary) indicates. Not so for red (diamond) and blue (circle) curves. Note a near perfect reflection symmetry between these curves. This is another manifestation of warming. The 1954 and the 2011 population maxima occur at ranks 3 and 63, respectively.  Hence, the statistically essential information for these years is stored in  intermediate ranks (see Fig. \ref{fig:1abcd} caption for further numerical illustration). On the other hand, high occupation of mid-rank, say rank 32, although significant, does not convey as much information about a warming trend as the high occupation of extreme, or near extreme, ranks. 

Based on the above discussion, the notion of a warming signature emerges, characterized by the over-population of the lowest ranks in early years, i.e.\ red values in upper left and lower right corners, with the blue values predominant in the other two corners.  But why limit one's attention to only symmetric partition of $P$ into four quadrants? To that end, consider the general partitioning into (unequal) quadrants defined by the off-center cross-hairs in Fig. \ref{fig:2abc}(a) and focus on the excess of records over the expected mean in quadrants 2 and 4, and the corresponding deficit in quadrants 1 and 3. For each quadrant {\em pair} we take the ratio of actual to expected populations and then form the difference of these two ratios. This difference vanishes (on average) for a steady climate. For the data of Fig. \ref{fig:1abcd}, the value of this difference at the centered cross-hairs $(32,32)$ is $14083/11680 - 9277/11680 = 0.4115$ while a peak value of $0.4283$ occurs at $(35,34)$. When this partitioning is repeated with the cross-hairs traversing the entire grid, a new matrix is generated, denoted as $Q$, e.g. $Q_{32,32} = 0.4115$. To ensure the existence of the four quadrants, given that $P$ is $N \times N$, the difference of ratios is computed at $(N-1) \times (N-1)$ grid points.\footnote{The row index of $Q$ is a time-like coordinate. Its $N-1$ values lie at the midpoints of the original grid with $N$ points.} The mathematical implementation for the above construction of $Q$ is given by
\be
\label{eq:qdiscrete}
\begin{split}
Q_{j,k} &= \frac{n_T}{n_t}\, \left [ 
\frac{\sum_{m=1}^{j} \, \sum_{n=1}^{k} \,
P_{m,n} + \sum_{m=j+1}^{n_T} \, \sum_{n=k+1}^{n_T} \, P_{m,n}}{j \, k + (n_T-j)\, (n_T - k)} \right . \\
 &\left . -
\frac{\sum_{m=j+1}^{n_T} \, \sum_{n=1}^{k} \,
P_{m,n} + \sum_{m=1}^{j} \, \sum_{n=k+1}^{n_T} \, P_{m,n}}{j \, (n_T-k) + (n_T-j)\, k}\right ] \, ,
\end{split}
\ee
where $n_T$ is the number of years. This defines the discrete $Q$ transform of $P$.
Note that $ -2 \le Q_{j,k} \le 2$, i.e.\ $Q_{j,k}/2$ is the excess or deficit percentage for the $(j,k)$ partition of $P$. If $Q$ and $P$ are rearranged as vectors, (\ref{eq:qdiscrete}) can be viewed as ${\bf q} = M \, {\bf p} $, where the matrix $M$, augmented with the row and column sum constraints for $P$, is well-conditioned and admits a stable inversion for $P$ given $Q$. Hence $Q$ preserves, while reordering, the trend information stored in $P$ from the original temperature record.

For the weather station of Fig. \ref{fig:1abcd}, the corresponding $Q$ is shown in Fig. \ref{fig:2abc}(b).  The complete trend information is stored in the set of partitions of $P$ and hence in the elements of $Q$.  As illustrated above, positive elements of $Q$ arise from partitions with a warming bias. Thus, for a stationary climate, one anticipates no sign preference for elements of $Q$. This motivates us to consider $\langle Q \rangle$, the mean value of all matrix elements, defined as
\be
\langle Q \rangle \equiv \frac{1}{N^2} \sum_{i=1}^{N} \, 
\sum_{j=1}^N  \, Q_{i,j}\label{eq:meanq}
\ee
The angular brackets, from now on, denote the average over all matrix elements throughout this paper (as opposed to an ensemble average) so $\langle Q \rangle$ is a scalar, and it vanishes on average for a stationary random process.  

Natural variability induces fluctuations in $\langle Q \rangle$ about zero.  Once that probability distribution is characterized, one has a quantitative basis to decide whether a trend is actually present, as discussed below. Thus, we propose to quantify a trend by the linear function (temperature vs. time) whose slope is determined by annulling the mean value of $Q$.  In other words, a single adjustable parameter, the slope, is chosen to annul the average matrix element of $Q$. To do this, a candidate linear function of $T(t)$ is subtracted from the original time series (input data), row-by-row ranks recomputed, $P$ re-populated with revised values, the $Q$ transform applied and its mean $\langle Q \rangle$ computed. The scalar $\langle Q \rangle$ is a monotone function of the trial slope and always has a single zero crossing. 

To illustrate, we return to the data in Fig. \ref{fig:2abc}. Remarkably, $Q$ is positive definite, that is, positive for each and every partition of $P$ (each matrix element of $Q$). Thus, the warming signature is exceptionally strong. Moreover, as Fig. \ref{fig:2abc}(c) confirms, not only does annulling $\langle Q \rangle$ in the original record determine a unique linear trend, that trend is nearly indistinguishable from the LS fit. Similar close agreement between LS and $Q$ linear trends is found in most cases.  Nonetheless, while LS and $Q$ fits of temperature trend commonly agree to $0.05  \degree$C over periods of 50 years or more, a few larger discrepancies arise. These arise in cases with  large seasonal variation in variance, which we shortly explore. A systematic cause of smaller discrepancies is that LS regression of the annual mean does not distinguish between a few large excursions in daily low temperature vs.\ numerous small excursions whereas $Q$ is affected principally by the latter. Lastly, autocorrelation, common in temperature time series, can differentially affect the two.

The partitioning of temperature data in a $365 \times 64$ matrix may seem a necessary condition for linear regression with $Q$. Not so. Dropping one calendar day to obtain $364 \times 64$ points affords a wide number of factorizations. The set $n_T = [26, 28, 32, 52, 56, 64, 91, 104, 112, 128]$ serves to make the point. Before detrending, $\langle Q \rangle$ values for this set consist of seven approximately equal low values, one intermediate, and two high. The last pair are the original $n_T = 64$, and subharmonic, $n_T = 32$, which averages two years of temperatures at a time. The superharmonic, $n_T = 128$ averages every six months hence the signal has both a long term trend and a period two seasonality. Its initial  $\langle Q\rangle$ is intermediate. The remaining seven, incommensurate with seasonality, all have a very irregular mean signal, though one still marked by the same long term trend. Each factorized form was detrended with exactly the same slope. All of them simultaneously have $\langle Q \rangle$ reduced to noise level (or, translated back to temperatures, differences averaging about $\pm 0.01 \degree$C).  So the choice of binning causes no meaningful disagreement about the trend required to annul $\langle Q \rangle$ on the assumption of a linear long term signal. 

\begin{figure}[ht]
\centerline{\includegraphics[height=3in]{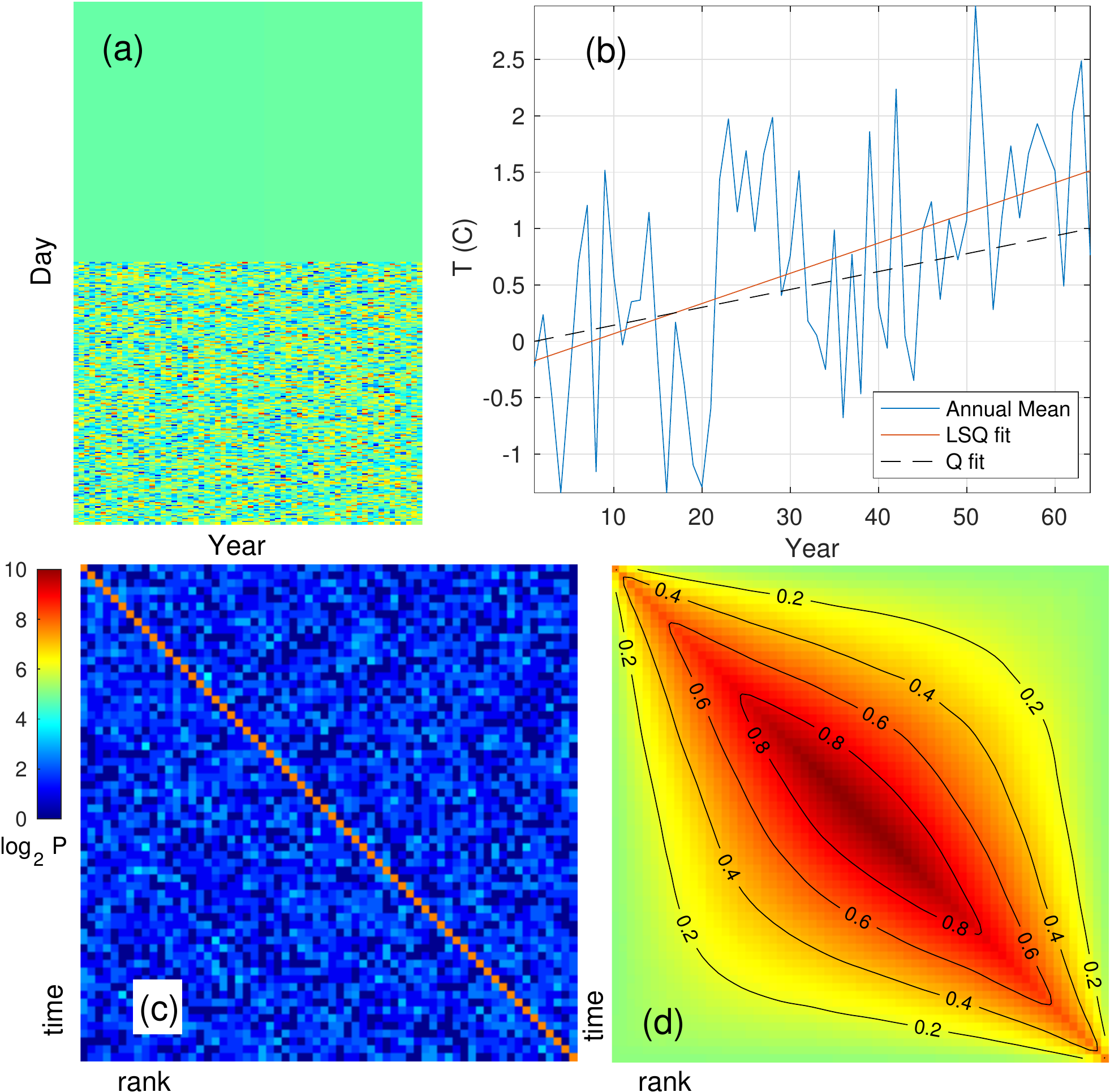}}
\caption{{\bf An example of superior $Q$ performance:} Synthetic data for daily low temperature on Planet X (see text). 
(a) the data matrix. For the first half of the year the temperatures are noise-free, only the trend of $1 \degree$C over 64 years is present. During the second half of the year large fluctuations are superimposed on the same trend; (b) A comparison of $Q$ and LS fits for this imaginary station. The $Q$ fit gives $1.0001 \degree$C. The LS fit is thrown off by the noise, giving $1.69 \degree$C with a confidence interval of $[1.06, 2.31]$. (c) The $P$ (log scale) matrix shows the reason for the disparity. The exact data for the first half of the year result in a diagonal population of entries while the second half of the year consists of randomly sprinkled entries; (d) $Q$ is hardly perturbed by noise; each partition sees a positive excess dominated by the diagonal.}\label{fig:toy}
\end{figure}

Note that the algorithm of obtaining the linear trend with $Q$ is objective in the sense that a robot can be programmed to detrend the temperature data by simply annulling $\langle Q \rangle$. A skeptical reader might wonder about extracting a dimensional quantitative trend in degrees/decade from the {\em dimensionless} rank input only.  In fact, it is signal and noise that together conspire to give $Q$ the quantitative information needed because ranks are scrambled by the noise indiscriminately while the signal affects them systematically. The key relation here is a proportionality constant that relates a dimensional change in slope to the dimensionless change induced in $\langle Q \rangle$ {\em for a specified noise field}. Unlike $Q$ itself, that constant does depend upon the exact distribution. We revisit this point at the close of Section \ref{sec:extension}, where an error estimate for slope is derived. 

As we shall see, the rank-order transform $Q$ reveals the entire form of a signal and not just the linear trend, i.e.\ there is information in the residual $Q$ shown in Fig. \ref{fig:2abc}(d). One does not generally expect $Q$ and least square fits to agree at all orders, particularly as ordinary least square fits are influenced by outliers, while $Q$ is not, for which see the treatment of heavy tails in Section \ref{sec:heavytail}, where empirically weighted least square fits can work up only to a point, while $Q$ performs well without need of such measures. Note also that a monotone deformation of temperature data (e.g., a logarithmic one) affects the least squares fit but not $Q$. 

To illustrate some remarkable properties of the $Q$ transform, we consider a highly idealized synthetic data set both because the true answer is known, hence $Q$ and LS errors can be assessed quantitatively, and because the idealization makes transparent the cause of the difference in comparative performance. Motivated by the data for Bethel Airport, AK where $Q$ and (unweighted) LS trends for 1951-2014 differ by $0.72 \degree$C, we consider the daily low temperature on Planet X, where the climate is so equable for the first half of the year as to have no variability in temperatures but solely a trend of $1 \degree$C over 64 years. In contrast, during the second half of the year the same trend is overlain with large variance. Fig. 3(b) shows the corresponding $Q$ and LS fits: $1.0001 \degree$C and $1.69 \degree$C, respectively. Clearly, the LS fit is thrown off by the abrupt noise. Fig. 3(c) depicts the $P$ matrix (note the log scale), revealing the reason for the divergent estimates. The exact data for the first half of the year result in a perfectly diagonal population of entries while the second half of the year consists of nearly randomly distributed entries. Fig. 3(d) shows the resulting $Q$ is resistant to noise; each partition sees a positive excess strongly dominated by the diagonal while the random entries largely average out.  Hence, detrending this $Q$, (see  (\ref{eq:idealQ}) for an exact expression in the limit of zero noise), effectively yields an exact result. In real data, all cases of large discrepancies in trend estimates between $Q$ and LS occur in locations that experience large excursions in seasonal variance. Conversely, $Q$ and LS linear trends for stations with minimal variance excursions commonly agree within the previously indicated $0.05 \degree$C per $50$ years. 

\section{Simple Analytic Approximations for $Q$\label{sec:Qapprox}}

Towards gaining an intuitive sense for $Q$ we introduce here a continuous version of $Q_{j,k}$, denoted as $q(x,y)$ and similarly for $P$. For simplicity, the domain of each is taken as $[-1,1] \times [-1,1]$. Then
\begin{equation}
\begin{split}
&q(x,y) = 
\frac{1}{2 - 2\, x\, y}\, 
\left [ \int_{-1}^x dx' \int_{y}^1  dy' \, p(x',y') + \right. \\
    &\qquad \left.   \int_x^1  dx' \int_{-1}^y dy' \, p(x',y') \right ]
        \\
&- \frac{1}{2 + 2\, x\, y} \, 
\left [ \int_{-1}^x dx' \int_{-1}^y  dy'\, p(x',y') + \right.\\
    &\qquad \left. \int_x^1  dx' \int_y^1 dy' \, p(x',y') \right ]\, , 
        \label{eq:qcont}
        \end{split}
\end{equation}
and we require $p(x,y)$ to satisfy the homogeneous constraints
\be\label{eq:pcon}
\int_{-1}^1 dx \,  p(x,y) = 0,\qquad 
\int_{-1}^1 dy \, p(x,y) = 0\, .
\ee
Making use of the latter constraints (\ref{eq:qcont}) can be simplified to
\begin{equation}\label{eq:qsimp}
\begin{split}
q(x,y) = \frac{1}{1- x^2\, y^2} \, 
&\left [ \int_{-1}^x dx' \int_{y}^1  dy'\, p(x',y') + \right . \\ 
  & \left . \int_x^1  dx' \int_{-1}^y dy' \, p(x',y') \right ]\, .
    \end{split}
\end{equation}
The inversion yields
\be
p(x,y) = \frac{1}{2} \, \frac{\partial^2}{\partial x\, \partial y} \,\left [ 
(1 - x^2\, y^2) \, q(x,y)\, \right ] .\label{eq:pinv}
\ee
Alternatively, we can write (\ref{eq:qcont}) in the form of a two-dimensional convolution as 
\be 
\begin{split}
q(x,y) &= \frac{1}{1 - x^2\, y^2} \, \int_{-1}^1\, \int_{-1}^1\, \left [ H(x-x')\, H(y'-y) \right . \\
&\left . + H(x'-x)\, H(y-y')\right ] \, p(x',y')\,dx'\, dy'\, ,
\end{split}
\ee
where $H$ denotes the Heaviside function. 

The simplest possible algebraic form that satisfies (\ref{eq:pcon}) is $p(x,y) = - x\, y$ and we choose the sign to reflect an excess in second and fourth quadrants and deficit in first and third, that is, a warming signal. From these assumptions results
\be\label{eq:lowmode}
q(x,y) = \frac{1}{2} \, \frac{(1-x^2)\, (1-y^2)}{1 -  x^2\, y^2}\, ,
\ee
with a mean value of $ {\pi^2}/{8} - 1 \approx 0.2337$ and root-mean-square value of $ \sqrt{1-3\, \pi^2/32}\approx 0.2733$.
While the issue of normalization has been bypassed, this simple {\it ansatz} for $p(x,y)$ is an excellent means to anticipate the {\em form} of a ubiquitous pattern in $Q$ both for real data at numerous sites with warming, and the dominant mode of $Q$ in a PCA decomposition, even for realistic correlated temperature fluctuations in a stationary climate, typically accounting for 25\% of the variance in $Q$. While (\ref{eq:lowmode}) reflects the form of $Q$ for a wide range of SNR, the limiting form for zero noise is a diagonal matrix for $P$. Translated to the continuous form, this results in 
\be\label{eq:idealQ}
\begin{split}
q(x,y) = &\frac{2}{x^2\, y^2 -1} \, 
\left [ 1 + x\, y - (x-1)\, H(x-1)\right . \\ 
&\left . -(x+1)\, H(x+1)
 +2\, (x-y) \, H(x-y)\right . \\
 &\left . +(y+1)\, H(-y-1)-(1-y)\, H(1-y) \right ]
\end{split}
\ee
whose diamond-shaped contours are those seen in Fig. \ref{fig:toy}. In this special case the formula above, if sampled on the unit interval at a spacing of $\Delta x = 2/n_T$ with endpoints excluded, is identical to the discrete result for $n_T$, regardless of the value of $n_t$. Appendix \ref{sec:breaks} examines breaks in a series, based on this continuous approximation.

\section{Metrics of $Q$, their statistical distributions and asymptotics}\label{sec:stats}

Reduction to a $P$ matrix is the basis for the $Q$ transform, and the exact general result for the equilibrium form of $P$ for a given signal in the presence of uncorrelated noise can be obtained. This result is essential for deriving error bounds. However, because of numerical complexity for realistic arguments, and the need for development of its asymptotic expansion, we defer that discussion to Appendix \ref{sec:asympt}. 

Here we extend our approach that began by consideration of $\langle Q\rangle$ in Section \ref{sec:deriveQ}. We aim to characterize the standard deviation for $\langle Q\rangle$ for iid noise. To this end we find an asymptotic expansion that clarifies parametric dependencies. Deeper meaning of such benchmarking emerges in the next section. 

In Section \ref{sec:deriveQ} we proposed that a linear trend can be determined by setting the average matrix element $\langle Q \rangle = 0$. Such a trend is a combination of a long term signal plus some contribution from natural variability. Given but a single realization, one cannot disentangle these two. However, knowing the distribution of $\langle Q \rangle$, one can set bounds on the contribution from natural variability to within any desired confidence level.  For iid noise, the quantity $\langle Q \rangle$ follows a normal distribution and the standard deviation of $\langle Q \rangle$ can be characterized in general terms.  Considering the disparate influence of $n_t$ and $n_T$ on that result, one expects the dependence on the former to be the same as that for a sum of $n_t$ normal variables, namely $n_t^{-1/2}$. It is plausible that an asymptotic expansion of $\sigma_{\langle Q \rangle}$ in $n_T$ has the same {\em leading order} dependence, succeeded by an ordered progression of higher order corrections. Numerical experiment at varying $n_T$ and $n_t$ with $6 \times 10^5$ realizations each time yields the following approximation in such a form:
\be\label{eq:meanq_asy}
\sigma_{\langle Q \rangle} \sim \frac{ 0.7131}{{n_t}^{1/2}}\,
\left [  \frac{1}{{n_T}^{1/2}} - \frac{0.2299} 
{{n_T}} +\frac{3.3026}{n_T^{3/2}} + {\cal O}( \frac{1}{{n_T}^{2} })\right ]\, .
\ee
(The coefficients above are sensitive to errors in computed estimates of $\sigma_{\langle Q \rangle}$.) As $Q$ is an ordinal method, asymptotic results such as (\ref{eq:meanq_asy}), and also (\ref{eq:qrms_asy}) below, are distribution-independent for white noise. The form above can be motivated by comparison to the derivation for a related expansion (see Appendix \ref{sec:asympt}). A sample run with 5000 trials using iid normal random variables, $n_t =365$, and $n_T =50$ gave $\sigma_{\langle Q \rangle} = 0.005456$ compared to the expected result from (\ref{eq:meanq_asy}) of $0.0054556$. Normalizing values of $\langle Q \rangle $ with the sample standard deviation yielded a distribution that passed the Kolmogorov-Smirnov test for normality at the $5$\% significance level with an asymptotic $p$-value of $0.035$.

%For that specific case one can still use the iid white noise baseline by first removing the seasonal signal from the data in Figure \ref{fig:1abcd}(a) and then shuffling the elements in each column, leaving a row-by-row delta-correlated time series. But more generally

Beyond linear trends, $Q$ may reveal a general nonlinear signal and a suitable second benchmark is then the root-mean-square (rms) value of $Q$ whose distribution must be characterized. The rms average  $\langle Q \rangle$ is given by: 
\be
Q_{rms} \equiv \sqrt {\langle Q^2 \rangle} =   \frac{1}{(n_T-1)} \left [ 
\sum_{j=1}^{n_T-1}\, \sum_{k=1}^{n_T-1} \, Q_{j,k}^2 \right ]^{1/2} \, .
\ee
The pair of mean and rms values of $Q$ have the great advantage that they are readily computed, especially the first for which there is a fast explicit algorithm given in Appendix \ref{sec:asympt}. (There is also a fast ${\cal O}(n_T^2)$ algorithm for $Q$ itself given $P$.)

% not sure here ${\cal O}(2 \, n_T^3)$ 

\begin{figure}
 \centerline{\includegraphics[height=2.75in]{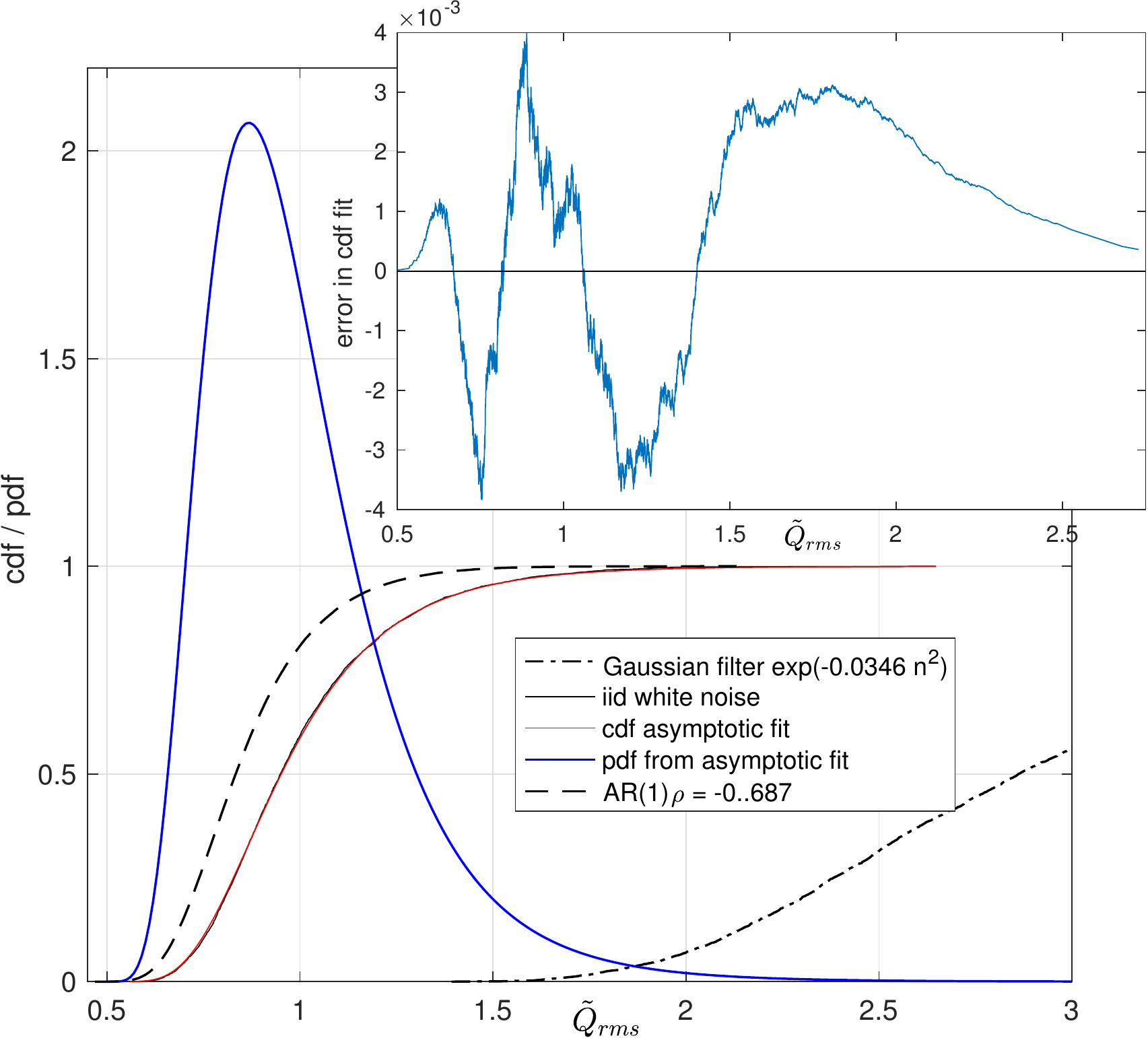}}
\caption{{\bf Distribution of $Q_{rms}$:} A baseline cdf of $Q_{rms}$ for iid noise and the asymptotic fit given at (\ref{eq:qrms_cdf}) (and corresponding pdf). Fit error is shown in the inset figure. Two comparison distributions show the effect of autocorrelation. The first (dashed line), with a narrower scale than the standard (solid line), is generated by an AR(1) model with $\rho = -0.68716$, which has a strong {\em negative} correlation at one time lag. The second, broader (dash-dot), is generated by iid noise convolved with a Gaussian filter. The latter two coincide exactly with the standard under a linear remapping of the abscissa. From (\ref{eq:qrms_asy}) this indicates the effect of autocorrelation amounts to a change in the effective $n_t$. Asymmetry of the distribution is clear in the pdf based on the asymptotic fit.}\label{fig:qrms}
\end{figure}

The quantity $Q_{rms}$ is observed to obey a generalized $\chi$-distribution and collapses to a single curve as a function of the normalized variable $ \tilde Q_{rms} \equiv Q_{rms} / \langle Q_{rms} \rangle$, and where a similar asymptotic expansion holds, namely:
\be\label{eq:qrms_asy}
\langle Q_{rms} \rangle \sim  \frac{1.3725}{{n_t}^{1/2}}
\, \left [ \frac{1}{{n_T}^{1/2}} + \frac{0.0293}{n_T} + \frac{1.3577}{{n_T}^{3/2}}
+ {\cal O}( \frac{1}{{n_T}^{2} }) \right ] \, .
\ee
An empirical expression for the cumulative distribution function (cdf) with uniform error $<$ $0.004$ can be written in terms of the incomplete gamma function\footnote{The incomplete gamma function used here is  8.2.2 of the {\sl Digital Library of Mathematical Functions}.} as
\be\label{eq:qrms_cdf}
\begin{split}
{\textrm{cdf}} (\tilde Q_{rms}) \approx 1 - &\frac{\Gamma(9.6070,13.6038\, \ln x + 9.9521)}{\Gamma(9.6070)}\\
& \qquad (x > 0.481) \, .
\end{split}
\ee

One must qualify the use of results like  (\ref{eq:meanq_asy}) and (\ref{eq:qrms_asy}) when the ambient noise is other than iid (white) noise. One common factor is autocorrelation. For example, it is a matter of common experience that weather has a persistence, typically 3 to 4 days. With the temperature data running vertically in the data matrix of Fig. \ref{fig:1abcd}(a), one has a resulting correlation between successive rows in that data matrix.  For correlated identically distributed variables arranged in this fashion it remains true that $\langle Q \rangle$ follows a normal distribution, but the coefficients in (\ref{eq:meanq_asy}) depend on the specific autocorrelation.  

For $Q_{rms}$ not only the coefficients change but the generalized $\chi$-distribution itself alters as seen in Fig. \ref{fig:qrms} where two examples make our point. The more conventional case is provided by convolving a Gaussian white noise sequence with a Gaussian filter of the form $\exp(-0.0346\, (n-n')^2)$. As in Fig. \ref{fig:1abcd}(a), the data are stacked vertically in the input matrix to $P$, hence successive rows are correlated. The resulting distribution of $Q_{rms}$ (dash-dot) is observed to be  broader. A second example, with a narrower distribution, is an AR(1) model with $\rho = -0.68716$, whose autocorrelation function has a pronounced dip of $-0.7$ at one time lag. The cdf for the standard reference $\tilde Q_{rms}$ (solid black), along with its asymptotic fit (\ref{eq:qrms_cdf}), lies between the other two. The difference between empirical and asymptotic results for iid noise is shown in the inset figure, and also the pdf that follows from the asymptotic form (\ref{eq:qrms_cdf}).

All three cdf curves are scaled by the same iid noise value for $\langle Q_{rms} \rangle$. For these two correlated examples, a linear remapping of the form $ \tilde Q_{rms} \to \alpha \, \tilde Q_{rms} + \beta$ gives a curve fairly close to the original iid distribution. The parameters that achieve this are $(\alpha = 0.2574, \beta = 0.0046$) for the Gaussian filter, and  $(\alpha = 1.3138, \beta = -0.1423)$ for the AR(1) model. The first of these, a shrinking of scale, can be thought of as a decrease in the effective number of independent samples $n_t$ \cite{koivunen1999feasibility}. That the second comparison distribution is narrower is attributable to the negative correlation, which disrupts, rather than reinforces, the tendency for transient nonstationarity. We draw upon this dynamic to great effect in Section \ref{sec:logistic}, where we consider chaotic series generated by the logistic map, also generally characterized by negative correlation.

% NOTE TRUE ANY MORE, CAN'T MATCH EXACTLY: In conclusion, provided that autocorrelation is accounted for with a suitable linear map of the independent variable for $Q_{rms}$, one can regard (\ref{eq:qrms_cdf}) as universal for {\em all} identically distributed random variables, correlated or not.

\section{Sample variability projected on the rank-order $Q$-plane characterizes stationary random processes. \label{sec:svd}}

The data shown in Fig. \ref{fig:2abc}(c) exhibit an unmistakable linear trend. Yet, at least in principle, natural variability of a truly stationary climate could create such a trend. While strict stationarity is a theoretical property of a random process, finite samples (even large ones) never appear purely random and exactly stationary. Finite samples exhibit transient trends and the likelihood of such trends depends on the specific stationary process. But, while spurious trends in sample mean and variance can be a hindrance for deterministic signal detection, one can turn this around and use these same calculated trend likelihoods to characterize (or ``fingerprint'') specific stationary stochastic processes. 

As we demonstrate below, the ``lifting'' of  a one-dimensional time series to the two-dimensional space of rank-order via the $Q$ transform enables an application of group theory, delivering a universal characterization of transient  trends for arbitrary stationary stochastic processes and sample sizes.  In particular, encouraged by anonymous reviewers, we pay special attention to two models: the Ornstein-Uhlenbeck process and the logistic map, the latter explored further from the perspective of deterministic chaos in Section \ref{sec:logistic}. 

\begin{figure}
\centerline{\includegraphics[height=1.5in]{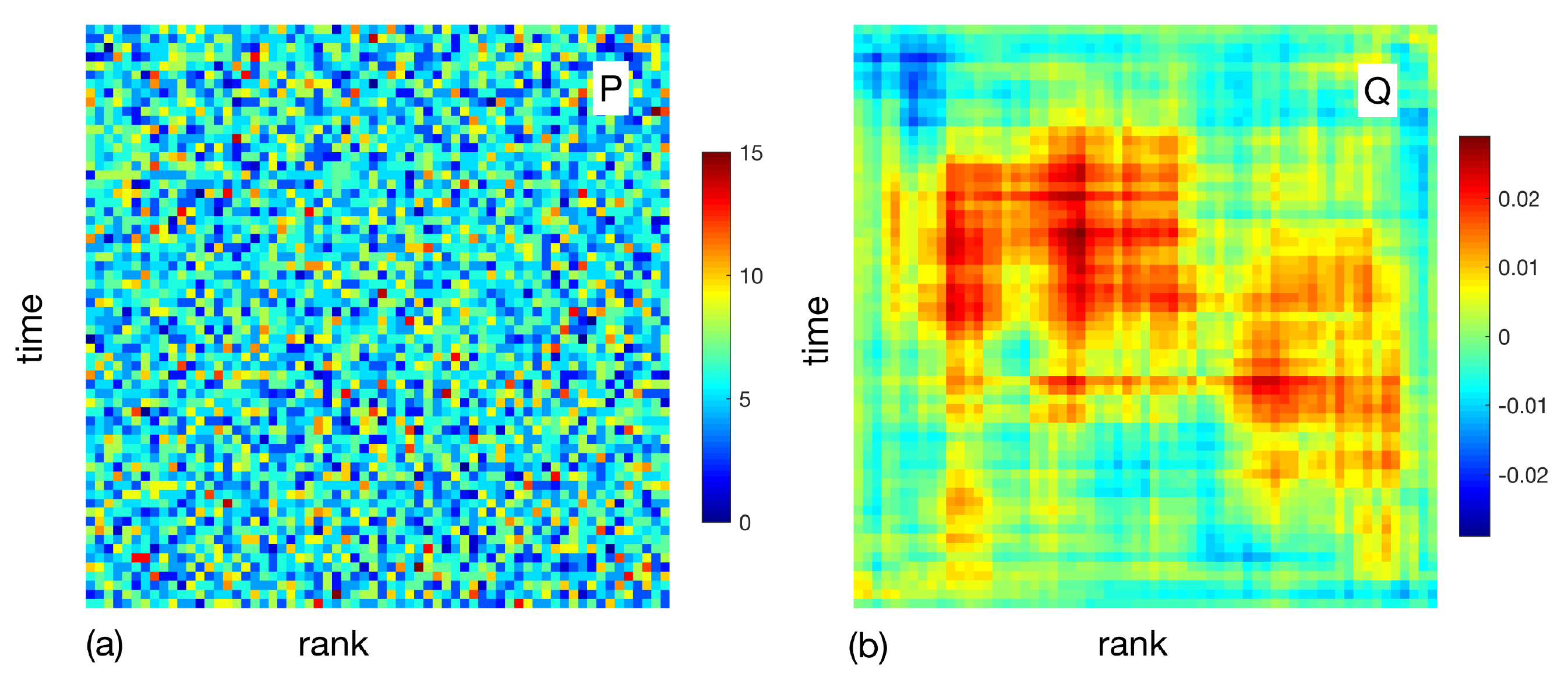}}
\caption{{\bf Contrast between $Q$ and $P$ representations:} (a) The fine speckle from a single realization of $P$ from the stationary iid Monte Carlo simulation, as described in text. (b) The corresponding spatial coherence in $Q$ (for any noise pdf). The average population per pixel on the left is $(365/64) \approx 5.7$.  Whereas $P$ is finely speckled, $Q$ exhibits a spatial structure, in this realization associated with a warming trend. Structures become more pronounced at the ensemble level (see Fig. \ref{fig:gsfig}).}\label{fig:FIG4}
\end{figure}

Towards the complete characterization of transient trends, we begin with the iid (stationary, $\delta$-correlated or white) noise, which is the featureless ``standard candle'' of stochastic processes. Because the $Q$ transform is ordinal, there is no need to limit our development to a Gaussian distribution; all white noise distributions are equivalent. The featureless spectrum of white noise suggests absence of features in any representation. Indeed, this featureless quality is so at the level of raw input data and remains true for $P$ matrix, e.g., see Fig. \ref{fig:FIG4}(a), devoid of apparent structure, appearing as salt-and-pepper noise. In fact, as all ranks have equal rights, ensemble-averaged $P$ tends to the perfect uniformity (constant $P$) for, not just iid, but more generally to all {\em independent} stationary processes because of the reshuffling argument (see Section \ref{sec:deriveQ}). This limit also holds for correlated (and hence, shuffling-breaking) stationary processes, aside from slight effects at the corners (see Appendix \ref{sec:exactP}).  

\begin{figure*}
\centerline{\includegraphics[height=3in]{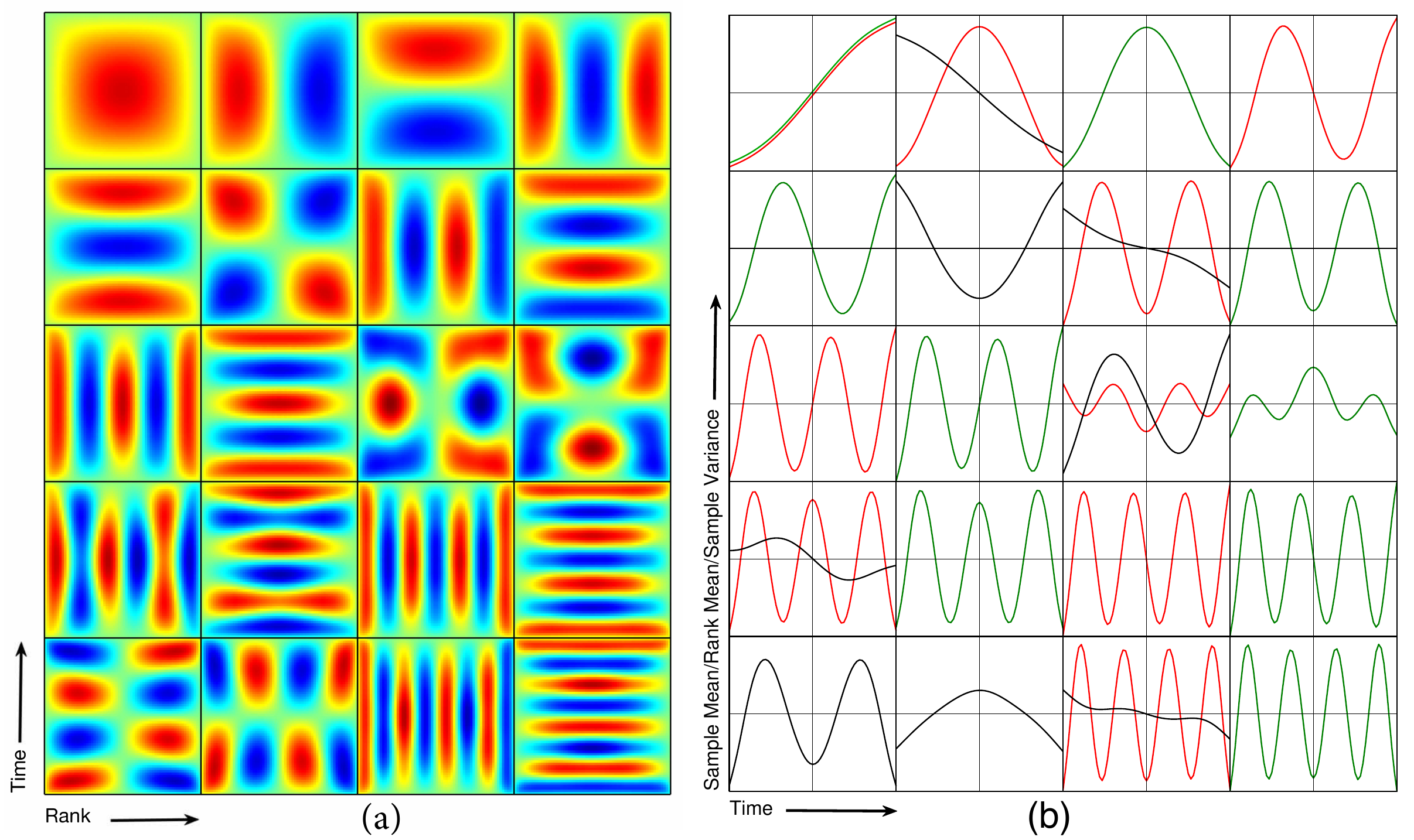}} 
  \caption{{\bf Universal modal decomposition of iid noise in the $Q$ representation:} a complete characterization of finite sample nonstationarity. (a) Merged PCA modes $\psi_k$ for $ k=1,\ldots 20$ from all five group projections, ordered (left to right) by decreasing singular value. (b) Transient nonstationary pattern of the corresponding time series: data mean $\delta \mu_k$ (green), data variance $\delta \sigma_k$ (black), rank mean $\delta r_k$ (red).}
  \label{fig:gsfig}
\end{figure*}

In contrast, the ensemble average of the $Q$-transformed (distribution-invariant) white noise in the rank-order plane (hereafter dubbed $\pi$ noise [34]) is {\em not} uniform and even at a single realization level, deviates greatly from the salt-and-pepper noise, as illustrated by the patchiness (structure) in Fig. \ref{fig:FIG4}(b). We take advantage of such structure and decompose it in terms of dominant modes (planforms), linking these planforms to the types of transient patterns in time (see Fig. \ref{fig:gsfig}). 

\subsection*{Group-based algorithm for the standard ``etalon''}

The desired correspondence between the planforms of $Q$ and specific features in the generating time series emerges from an examination of symmetries and associated groups. Group character is central in the rank-order plane, e.g., time-reversal symmetry means the ensemble average of $P$ is invariant under a left-right flip.  Just as any 1-D function $f(x)$ can be written as the sum of even and odd terms, $1/2[f(x) + f(-x)] + 1/2[f(x)-f(-x)]$, an arbitrary function in $n$ dimensions has a unique, orthogonal group decomposition in $n! + 2^n -1$ terms (two terms for $n=1$). For $n=2$ the five term expansion assumes the form
\be\label{eq:qdecomp}
q(x,y) = q^{(D_4)} + q^{(D_2)} + q_x^{(C_1)} + q_y^{(C_1)} + q^{(R_2)} \, ,
\ee
where
\begin{align*}
q^{(D_4)} &= [ q(x, y) + q(-x, y) + q(x, -y) + q(-x, -y) \\
& + q(y, x) + q(-y, x) + q(y, -x) + q(-y, -x) ] /8\\
q^{(D_2)} &= [ q(x, y) + q(-x, y) + q(x, -y) + q(-x, -y) \\
 & - q(y, x) - q(-y, x) - q(y, -x) - q(-y, -x) ] /8
\end{align*}
\begin{align*}
q^{(C_1)}_x &= [ q(x, y) + q(-x, y) - q(x, -y) - q(-x, -y) ] /4\\
q^{(C_1)}_y &= [ q(x, y) - q(-x, y) + q(x, -y) - q(-x, -y) ] /4\\
q^{(R_2)} &= [ q(x, y) - q(-x, y) - q(x, -y) + q(-x, -y) ] /4\, .
\end{align*}
Here $D_n$ denotes the dihedral group, $C_n$ the reflection group, and $R_n$ the rotation group, with the third and fourth components on the right in (\ref{eq:qdecomp}) representing reflections about the $x$ and $y$ axes respectively. The applications of this expansion appear manifold, including an exploration of wallpaper groups as in \cite{Verberck2012}\footnote{As commented in [20], ``A more formal approach for deriving minimal symmetry-adapted functions for the wallpaper groups involves group theory; each wallpaper group should be decomposed into irreducible representations.'' We note one result in this area, that the wallpaper group $p6mm$ of graphene has a point group expansion from (\ref{eq:qdecomp}) in $D_4$ and $D_2$ only.}. The first term is the only one in the decomposition with (in general) a nonzero mean value when integrated over the domain; all others vanish identically by anti-symmetry.\footnote{In higher dimension, the first term of this decomposition has symmetry $B_n$, the hyperoctahedral group.}

% Note [33] (currently) here is hardwired. We need a forward reference to a FOOTNOTE, not a bibliographic entry. Using \label inside the footnote does not work. 

The expansion (\ref{eq:qdecomp}) can be applied in discrete form to the square matrix $Q$ for each realization\footnote{Among the properties of (\ref{eq:qdecomp}) yet to be explored, for a square matrix populated by iid normal entries, the ensemble variances are evidently $\sigma^2_{q_{(x,y)}^{(C_1)}}  = 1/4$, $\sigma^2_{q^{(D_2)}} = 1/8$ as the matrix dimension tends to infinity, with the remainder apportioned in enigmatic proportion between $q^{(D_4)}$ and $q^{(R_2)}$.}, yielding five ensembles; one per group. Each of these ensembles is then characterized by principal component analysis (PCA). This expansion is driven by data (hence Lorenz's term ``empirical orthogonal functions'' \cite{Lorenz1956}), rather than pre-selected, as in a generalized harmonic analysis of noise. PCA is designed to decorrelate the signal by projecting the data onto orthogonal axes. Here it decomposes $\pi$-noise variability in the $Q$ group representation with modes in order of decreasing contribution to variance ($\sigma^2_Q$, a quadratic metric) of each ensemble.  

\begin{figure}
\centerline{\includegraphics[height=4in]{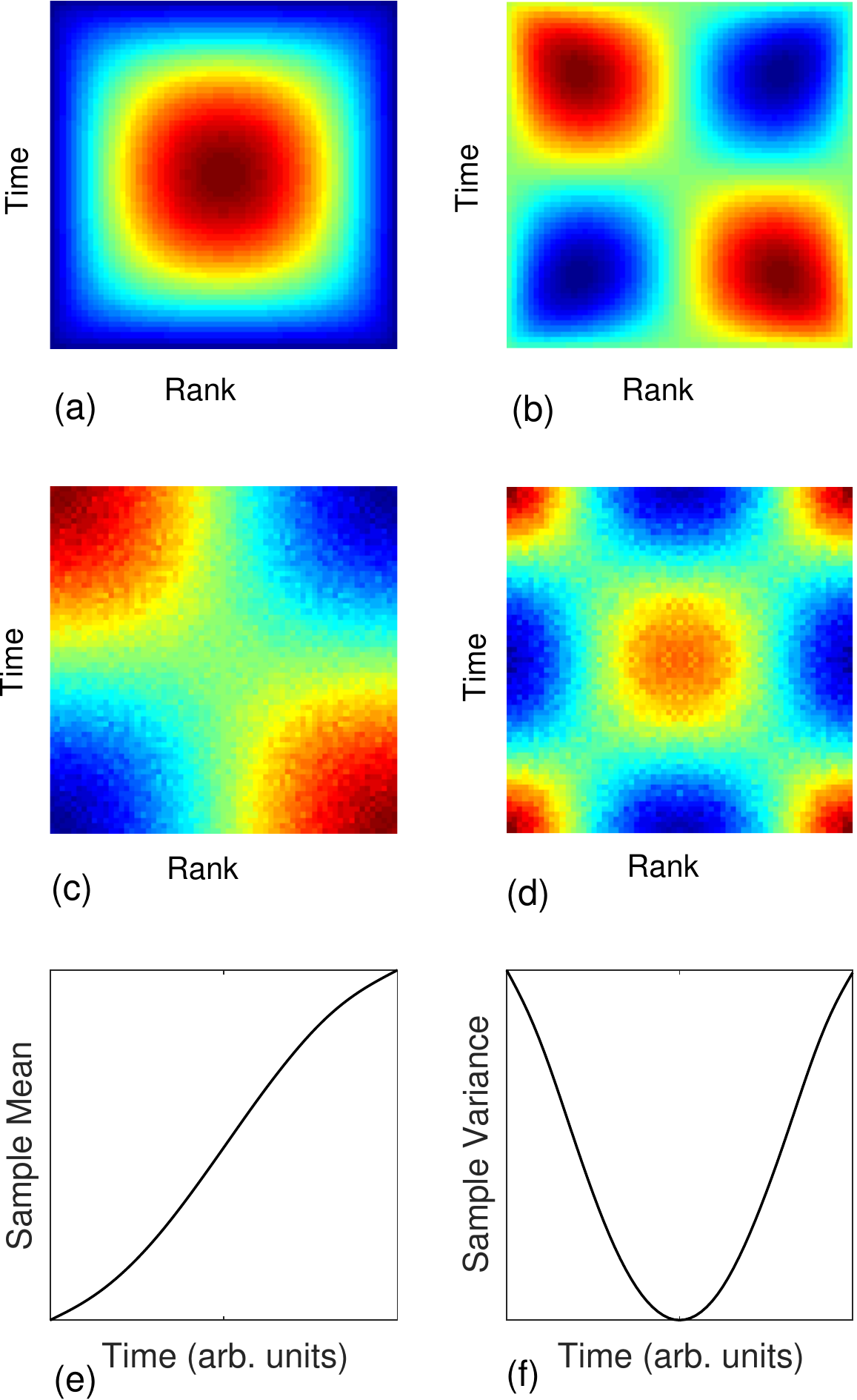}}
\caption{{\bf Examples of Modes 1 and 6:} (a) $\psi_1$ with $D_4$ symmetry; (b) $\psi_6$ with $R_2$ symmetry (both from from Fig. \ref{fig:gsfig}(a)); (c) inverting $\psi_1$ for $P_1$ from (\ref{eq:qdiscrete}) gives a result with $R_2$ symmetry; (d) same procedure for $P_6$ has $D_4$ symmetry. Similarly (e) conditionally averaged sample mean of the data for $\psi_1$ (averaged over $n_t$ trials), and resembles the Lugano temperature time series of section II. In (f) the vase-like profile in sample variance also from conditional sampling represents over-population of the four corners of $P_6$. (Both (e) and (f) from Fig. \ref{fig:gsfig}(b).)}\label{fig:casestudy}
\end{figure}

For numerical implementation, PCA is evaluated by singular value decomposition (Matlab routine {\tt svd}). We used an ensemble of $10^5$ realizations populated by iid normal random variables of zero mean and unit variance though the ordinal results depend on neither choice, even from row to row. For $\delta$-correlated noise, the lowest modes from the PCA decompositions of the resulting $Q$ group ensembles rapidly approach their limiting forms as a function of $n_T$, with the highest retained mode $\psi_{20}$ determining the needed grid resolution. We aim for well resolved structure, not just meeting the Nyquist limit. As a test of this, spline interpolation of $\psi_1$ for $n_T=65$ onto the coarser mesh of $\psi_1$ for $n_T = 49$ gives a relative standard error for the mismatch of $3\times 10^{-3}$. The singular values (scaled by $n_T^{-1/2}$) exhibit a similar relative error. The choice of $n_T=65$ will thus suffice for most applications and so one need not repeatedly compute this etalon for different $n_T$ but rather can rely on interpolation.

When searching for signal in noise, $Q$ approaches a finite limiting form as $n_t^{-1/2}$. Here {\em there is no signal} and hence no structure which $Q$ attains with increasing $n_t$. Remarkably then, and quite in contrast with e.g.\ the temperature data for Lugano, the PCA results for iid noise with the minimum possible choice of $n_t = 2$ are indistinguishable from those for $n_t = 2048$. The reduction to  $n_t=2$ saves CPU time for both generating the random realizations and their initial processing to obtain $P$. 

For each symmetry group, the PCA modes have an ordered set of singular values. The collected set of all group PCA modes is then re-sorted by singular value, with the corresponding symmetry group noted for each. In this merged set one encounters repeated mode pairs of symmetry $(\psi_{D_4}, \psi_{D_2})$ and  $(\psi_{R_2}, \psi_{R_2})$. In both cases transient nonstationarity is more compactly represented by forming sum and difference modes, i.e.\ $( \psi_{D_4} \pm \psi_{D_2})/\sqrt{2}$, and similarly for the other pair. There are also unpaired modes of all three symmetries, particularly at higher order. But, of the first twenty modes, only two such exceptions occur: the first and sixth modes, to which we shortly turn.\\

\subsection*{Results: a universal characterization of transients for $\pi$ noise}

Fig. \ref{fig:gsfig}(a) shows a set of $x-y$ oriented planforms, indexed as $(j,k)$ denoting a total of $j$ extrema in the $x$ direction and $k$ in $y$. The case of $j= k$ corresponds to the above two unpaired modes $j=\{1,2\}$ while for $j \ne k$ we have pairs in the form of a matrix and its companion transpose. This then constitutes our ``etalon'' against which stochastic processes are to be compared.\footnote{The $x-y$ categorization used here breaks down at higher order when further bifurcations cause complex patterns but these are not of practical concern.}  
% end of etalon description

$\pi$-noise variability falls into three main categories: nonstationarity of the sample mean, $\delta\mu_k$; of sample variance, $\delta\sigma_k^2$; and departure from $\delta$-correlation, described by the autocorrelation function (ACF) for {\em stationary} random processes \cite{yaglom2004introduction}. For reasons of symmetry in the rank-order plane, we also consider nonstationarity of (sample) mean rank, $\delta r_k$. These curves are obtained by conditional sampling in a long Monte Carlo run. Each realization with a mode projection for $\psi_k$ exceeding the $2 \sigma$ level is captured. The means of the realizations thus isolated, mode-by-mode, are plotted in the matching tableaux of Fig. \ref{fig:gsfig}(b). These curves (time series) follow the group selection rules indicated in Table \ref{table:planform}.\footnote{For simplicity, we have omitted modes for spurious trends in variance of {\rm rank}, which would symmetrize the table with a complementary entry in the third line.}

\subsubsection*{Tutorial on Fig. \ref{fig:gsfig}: Case studies for modes 1 \& 6}

Although PCA modes for $P$ are of little use, each $Q$ mode $\psi_k$ can be inverted to discover its antecedent $P$.\footnote{The magnitude of $\psi_k$ is arbitrary when this is done but has an upper bound which, if exceeded, induces negative elements in the resulting $P$.} In Fig. \ref{fig:casestudy}(c,d) we show the $P$ precursors for two modes, $\psi_1$ and $\psi_6$, reproduced here from Fig. \ref{fig:gsfig}(a). These examples will demonstrate how separation of transient mean and transient variance arises from ``lifting'' to the rank-order plane. (Other similar separations are also seen in Fig. \ref{fig:gsfig}, e.g. modes $\psi_{17}$ and $\psi_{18}$, associated solely with variance.)  

The first mode $\psi_1$ is associated with an approximately linear trend in data, $\delta \mu_1(x)$.\footnote{$\delta \mu_1(x)$, with one inflection point, is not quite linear, hence an exactly linear trend in the data maps onto an {\em expansion} in odd modes, though dominated by the first. Note that $\delta\mu_k(x)$  and $\delta r_k(x)$ curves, each as a set, are not orthogonal, in contrast to the group PCA modes.} How can one see this intuitively? Here $P_1$ proves essential. Imagine a realization for which Fig. \ref{fig:casestudy}(e) is, by chance, the mean trend. Record lows (and generally lower ranks) are more likely to occur at early times and, conversely, record highs at later times.  Such an excess of record lows in the upper (early time) left (lower rank) corner of $P$ paints it red, and similarly for the lower right corner, paralleling the construction that lead to Fig. \ref{fig:2abc}. Thus, a linear trend of the mean yields $P_1$, odd in both its dimensions and corresponding an even/even $Q$ (consistent with the mixed derivative in (\ref{eq:pinv})). Not only linear, but anti-symmetric, trends lead in general to $R_2$ symmetry of $P$ and $D_4$ symmetry of $Q$.\footnote{The link of an antisymmetric trend and pure $D_4$ symmetry of $Q$ applies to stationary processes. (For iid white noise in particular it derives from (B1) perturbed about the vacuum state.) However, for finite SNR $P$ acquires lower symmetry components as well, e.g.\ $R_2$ in Figure 3(c). Nonetheless annulling $\protect\langle Q\protect\rangle$ remains valid. Also, for a measure of nonstationary mean of {\em any form}, but which excludes nonstationary variance, one can modify (11) by computing the rms value of the horizontal mean of $Q$.}

The second mode $\psi_6$ is paired with a roughly quadratic profile in variance. Again, by appealing to $P_6$, we can understand this relation by considering a realization with sample mean variance as in Fig. \ref{fig:casestudy}(f). Now {\em both} record highs and lows are more likely to occur at early and late times, thereby producing the red corner pattern of Fig. \ref{fig:casestudy}(d). Further, the over-population of middle ranks at intermediate times also paints the center of $P$ red. Then, because of the row and column sum constraints, necessarily all four middle edges must be under-populated (blue). A similar derivative argument applies for parity,  and a general statement is that symmetric trends in variance lead to $D_4$ symmetry of $P$ and $R_2$ symmetry of $Q$.

\bigskip\noindent
Returning now to Fig. \ref{fig:gsfig}(b), note the consecutive identical pairs of mean rank (red) and mean data (green),  that is $\delta \mu_2(x) = \delta r_3(x)$ for $\psi_2$ and $\psi_3$ respectively, and so on. Mode pair $(\psi_{11},\psi_{12} )$ marks a planform bifurcation from $j=0$ to $j=1$ in lines 3 and 4 of Table 1, with a more subtle relation to $\delta r_{11}$ and $\delta\mu_{12}$.

The only member of the odd/odd planform category here is $\psi_1$, but the notation in Table \ref{table:planform} anticipates presence of a higher planform $(3,3)$ also of $D_4$ symmetry. Mode 21 from the merged PCA expansion is that planform. The leading four PCA modes of this merged set account for nearly half the total variance while the asymptotic decay rate is $\sim n^{-\ln 2}$ \footnote{The appearance of $\ln 2$ here suggests that binary decision underlying ranking can play a role. This is reminiscent of $k\, T\,  \ln 2$ in the Landauer principle.}, in contrast to the ``whitish'' $\sim n^{-\varepsilon}$ for raw input or $P$.  

Transient trends in (sample) variance are plotted in black. Note how modes of $R_2$ symmetry (6,17,18) are associated with spurious trends in sample variance alone, just as modes of $D_4$ and $D_2$ symmetry are linked to odd order trend of {\em only} the sample mean. It is the $C_1$ pairs where odd order variance and even order mean are linked. 
  
The notation $\delta (r, \sigma)_k(x)$ reminds one that these modes are zero-mean fluctuations. But ensemble means from conditional sampling are {\em not} zero-mean. Rather, the conditionally sampled modes for {\em rank} all have mean $(1+n_T)/2$ and similarly the modes for variance have a mean equal to that for a sum of $n_t$ values of a random variable from the particular distribution used, here unity. The negative values in the plots for Fig. \ref{fig:gsfig}(b) then are relative to these means. Both rank and variance themselves remain positive definite. For graphical purposes only a single re-scaling was applied to all curves in Fig. \ref{fig:gsfig}(b), so their relative magnitudes can be compared directly.  

While results based on rank, as for any results from $Q$, are distribution-independent, transient {\em dimensional} fluctuations in mean and variance refer back to the raw data space and these necessarily reintroduce a dependence on the particular distribution in question. At issue is a constant of proportionality between, say, a given gradient in dimensional variables and the induced change in the dimensionless measure $\langle Q \rangle$. We treat this for the specific case of Gaussian noise later in Section \ref{sec:extension}, where we derive an explicit error estimate for $Q$-based linear regression. The theoretical framework for making that link is given in Appendix \ref{sec:asympt}.

Note that $\langle Q \rangle$ automatically annihilates all modes except those of groups $D_4$ and $D_2$. The latter group occurs in pairs. Each such mode pair $(\psi_k, \psi_{k+1})$ can be rotated {\em back} to the original basis by $(\psi_k \mp \psi_{k+1})/\sqrt{2}$. Only the recovered mode of $D_4$ symmetry then contributes to $\langle Q \rangle$. The second -- in which trends in rank and mean are {\em anti-correlated} -- vanishes identically in integral.

\begin{table}
\begin{center}
 \begin{tabular}{|c| c |c | c |} 
\hline
  Planform  & $Q$ sym   &  $\delta P$ sym & Null Projection\\
\hline\hline
$(2j+1, 2j+1)$  &  $D_4$  &  $R_2$  & $\delta\sigma^{(\pi)}_k(x) =0 $\\
\hline
$(2j+1, 2k+1)\quad j\ne k$ & $D_2$& $R_2$   &   $\delta\sigma^{(\pi)}_k(x)=0$ \\
\hline
$(2j+1, 2k)$  &   $C_1^{(y)}$  & $C_1^{(x)}$ &$\delta \mu_k(x) = 0$\\
\hline
$(2k,  2j+1)$ &   $C_1^{(x)}$  & $C_1^{(y)}$ &  $\delta r_k(x) = \delta \sigma_k(x) = 0 $\\
\hline
$(2j, 2k  )$  &   $R_2$  & $D_4 \pm D_2$ & $\delta r_k(x)= \delta \mu_k(x) = 0 $\\
\hline
\end{tabular}
\end{center}
\caption{\label{table:planform} 
Col. 1: Planform patterns for the modes $\psi_k$ in Fig. \ref{fig:gsfig}. Col. 2: Their symmetry group: dihedral group $D_n$, reflection group  $C_n$, and rotation group $R_n$. Col. 3: Symmetry group of the companion precursor $P$ field. Col. 4: Associated fluctuation fields that vanish identically.}
\end{table}

Anti-correlation is forbidden at lowest order; the only mode present already is of group $D_4$. Linear trend in rank must hence match trend in the data regardless of the loss of magnitude information. This is not obvious. One can try to construct a companion $Q$ mode, necessarily of group $D_2$, with rank and data linearly anti-correlated, e.g. $x^2-y^2$ in continuous form but inversion of any such form yields a $P$ of singular support, that is a set of measure zero for projections from the space of ranked white noise realizations. The problem is that one needs a form for $Q$ which vanishes on the boundaries but at the same time satisfies (in the continuous version)
\[
\frac{d}{dy}\, \int_{-1}^1 \, q(x,y)\, dx \sim y \quad \textrm{and} \quad
\frac{d}{dx}\, \int_{-1}^1 \, q(x,y)\, dy \sim -x\, .
\]
and this is evidently not possible.  

We can now give a precise statement of the meaning of annulling $\langle Q \rangle$: The initial data yield a nonzero $\langle Q \rangle$ from the sum of projections on $D_4$ modes only (subject to the second rotation noted above).\footnote{From the expansion in (\ref{eq:qdecomp}) for Lugano, $Q^{(D_4)}$ accounts for 99.2\% of the content of the $Q$ matrix in Fig. 2(b), the highest fraction for any station observed.} Adding a linear trend to the data modifies the contributions, principally from mode 1. The coefficient of that linear term is adjusted until the total sum from all $D_4$ terms vanishes. As explained in the discussion of Figure \ref{fig:casestudy}, this procedure is unaffected by nonstationary variance. The invariance of $Q$-derived trends of the mean with respect to variance thus holds unconditionally. 

%Moreover, as the group PCA basis set is complete, each of the (discrete) components of (\ref{eq:qdecomp}) has an expansion in modes $\psi_k$ of the same group and hence the selection rules of Table \ref{table:planform} also apply to (\ref{eq:qdecomp}). 

This is the crucial difference between least squares and $Q$. Least squares fits are strongly affected by non-stationary variance, as shown by our earlier toy model of Figure \ref{fig:toy}. One has then to resort to empirically determined weights to try to minimize this influence. For heavy-tailed noise however such weights prove ultimately ineffective, as we later document in Section \ref{sec:heavytail}. No such empirical machinery is needed for $Q$.

Note that if the goal is merely to obtain a trend by annulling $\langle Q \rangle$, then {\sl any} functional form with nonzero antisymmetric component will also project on the $D_4$ modes and hence determine a unique amplitude for that function. Annulling $\langle Q\rangle$, that is, does not confer any special status on a linear trend. Rather, that choice resides in the application and the onus is on the user to choose. 

A second moment of interest is $\langle Y \circ Q \rangle$ (where $\circ$ represents the Hadamard product). This selects for only the modes in group $C_1$ with parity $+ -$, which are raised to $D_2$ and parity $++$, and thus contribute in integral. This is the natural companion measure to detect even signals of nonstationary mean while $\langle Q \rangle$ detects odd.

Note the generality of these results: the $Q$ response to an actual signal of low SNR results from combining the components in Fig. \ref{fig:gsfig}(a) weighted by the expansion coefficients for that signal when expressed in terms of the complete set $\{ \delta\mu_k\}$ in \ref{fig:gsfig}(b). Hence, whether considering the transient sample mean of a stationary process, or the real mean of a non-stationary one, $Q$ detects them the same. The key distinction is that, for the case of $\pi$-noise, the standard deviation for each of these modes is universal and fixed and their means vanish; for a signal, the amplitudes are arbitrary and unknown in advance.

\section{Fingerprinting Stochastic Processes \label{subsec:fingerprint}}

\begin{figure*}
\centerline{\includegraphics[height=3in]{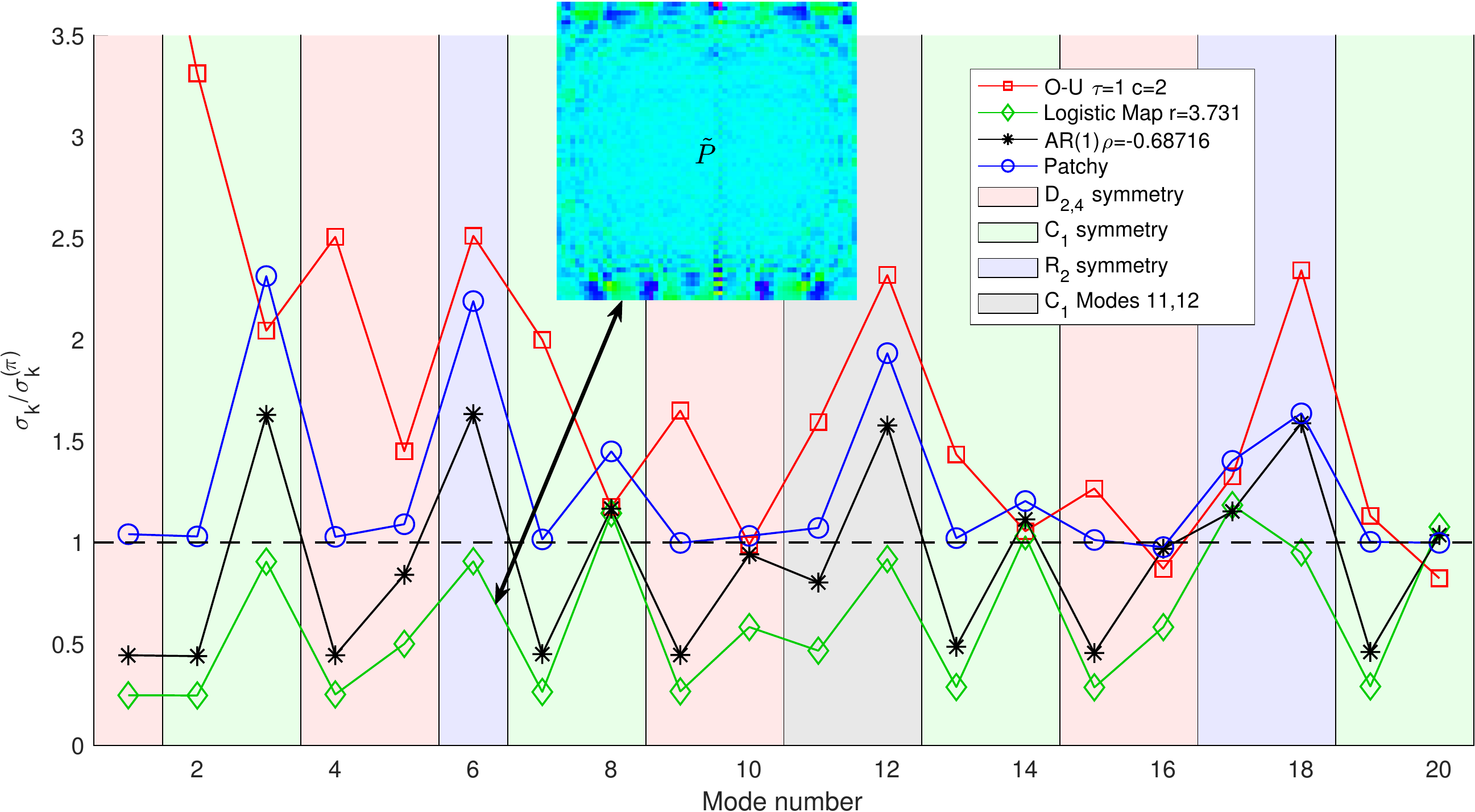}}
\caption{{\bf Stationary Stochastic Processes and Deterministic Chaos Characterized by Patterns of Rank Sampling Variability}. Y-axis: Standard deviation $\sigma$ of the modal coefficients normalized by the $\pi$-noise values. Three models are stationary stochastic processes: (i) the Ornstein-Uhlenbeck, (2) ``patchy'' $\delta$-correlated, (3) first-order autogressive (AR(1)). The fourth is the chaotic logistic map (examined further in Section \ref{sec:logistic}). For the parameters noted in the legend, the AR(1) and logistic model have essentially equal ACFs but distinct fingerprints. Both dip below the $\pi$-noise because of negative correlation at small lags, reducing the likelihood of a spurious trend. The inset shows $\tilde P$ (the ensemble mean $P$ absent its $D_4$ component) for the logistic map. For all independent stationary stochastic processes, $\tilde P = 0$ but for all deterministic processes, as here, $\tilde P \ne 0$. The patchy process illustrates that $\delta$-correlated processes can still have nontrivial fingerprints. The most telling feature of the O-U process is its greatly increased probability of spurious linear trends relative to $\pi$ noise as indicated by its offscale value for mode 1 of $4.97$.}\label{fig:fprint2m}
\end{figure*}

% previous figure caption
%Results of ensemble averages of $ 10^5$ realizations for each of the following random processes: O-U process with relaxation parameter $\tau =1$ and diffusion constant $c = 2$ (as in \cite{Lehle2011}), the auto-regression process AR(1), with $\varphi = -3/4$, the logistic map at $r = 3.8$, and the doubly stochastic ``patchy'' process (see text).  The two processes dipping below the $\pi$-noise are the ones with negative correlation at small lags (logistic map and AR(-0.75)), reducing the likelihood of a spurious trend, e.g., values $< 0.5$ for mode indicate reduced probability of a transient linear trend in rank relative to $\pi$ noise. The associated group PCA modes are color-coded by symmetry, e.g., light blue for $R_2$.

The group PCA decomposition yields the $\pi$-noise standard deviation for each of the first 20 modes, thus defining benchmarks. Stationary processes other than white noise will deviate in one or more of these measures, just as earlier observed in Section \ref{sec:stats} with the influence of autocorrelation on the distribution of $Q_{rms}$. Although group PCA components represent apparent non-stationarity, spontaneously arising in a finite sample of a random process, each standard deviation for the parent distribution of individual mode coefficients has an asymptotic expansion of the same general form as (\ref{eq:meanq_asy}) and (\ref{eq:qrms_asy}). Hence the suite of {\em ratios} of such quantities  (a ``fingerprint''), approaches a well-defined limit as $n_T \to \infty$. 

The four processes illustrated in Fig. \ref{fig:fprint2m} are: the Ornstein-Uhlenbeck (O-U) process with relaxation $\tau =1$ and $c = 2$ (as in \cite{Lehle2011}), the auto-regressive process AR(1) with $\varphi = -0.68761$ \cite{priestley1981,percival1993}, a model for patchiness consisting of white noise with the standard deviation for each successive group of 13 samples chosen from a uniform distribution on the interval $[0,1]$, and a chaotic series generated by the logistic map with $ r = 3.731$. 

This group-theoretic signature, consisting of the standard deviation for each mode normalized by the $\pi$-noise values, is one way to detect and/or classify a specific stationary stochastic process. It is a function of $n_T$ (but not $n_t$) just, as in the correlation theory of random processes, the ACF is a function of the number of time lags, $n_\tau$. But the fingerprint furnishes information over and above that available from the ACF. Distinct stochastic processes with nearly identical ACFs are shown in  Fig. \ref{fig:fprint2m}: (1) the $\delta$-correlated (like $\pi$ noise``) patchy'' process whose fingerprint oscillates about the $\pi$-noise standard; (2) the $AR(1)$ model and the logistic map with distinct fingerprints.

The largest departures from $\pi$-noise occur for the O-U process, with a long correlation, in contrast to the $\delta$-correlated patchy process. This fingerprint of O-U can be compared to the approach to stochastic signal detection in \cite{Lehle2011}, but with the further development in \cite{Lehle2017}, generalized there from a parametric to a non-parametric version based on higher moments of noisy data. Our method is also non-parametric, but deals only with rank and hence serves as a complementary approach to \cite{Lehle2017}.

Fingerprints of stochastic processes should be compared at the same $n_T$ (or $ n_T\, \Delta t$ in the continuous case). As $n_T$ attains a value several times the longest expected correlation, the fingerprint attains its asymptotic limit. For three of the four processes in Fig. \ref{fig:fprint2m}, $n_T = 65$ is well into that regime. However, the continuous O-U process has a much longer correlation time and, for a step size of $\Delta t = 0.01$, one would need $n_T$ of order $10^3$ to reach that limit. Its fingerprint at $n_T=65$, strongly dominated by the (offscale) peak mode 1, is nonetheless a perfectly fair point of comparison with any other stochastic process {\em at the same $n_T$}.\footnote{One case, not shown in Fig. \ref{fig:fprint2m}, yields no discernible departure from $\pi$-noise and that is the first billion decimal digits of $\pi$. This holds for strings of $1$, $2$, $3$, and $4$ digits after accounting for ties. Thus, $Q$ perceives digits of $\pi$ as iid noise, hence the name.}

\subsubsection*{Generality of results}

The suppression of apparent linear trends (mode 1) by both the logistic map and AR(1) in Fig. \ref{fig:fprint2m} evidently reflects an inhibiting effect of the negative correlation at one time lag in the ACF.  But the hallmark of true, rather than apparent sample, non-stationarity is the presence of structure in $P$, as for any deterministic signal, buried in noise or not.  This is in contrast to the constant (uniform) ensemble-averaged matrix $P$ that obtains for any stationary random process (but see Appendix \ref{sec:exactP} for a small caveat which explains the removal of the $D_4$ component of $P$ as in Fig. \ref{fig:casestudy}, hence the $\tilde P$ in Fig. \ref{fig:fprint2m}). Chaotic systems are deterministic and even the logistic map at $r=4$, commonly thought to be random, has a structured $\tilde P$. All chaotic systems exhibit intricate, and unique, ensemble-averaged patterns for $\tilde P$. One of the discoveries of this paper is that ranking within randomness differs inherently from ranking in chaos (as well as more orderly deterministic systems) as reflected in rank portraits (analogous to phase portraits), e.g., $\tilde P$ in Fig. \ref{fig:fprint2m}). 

% have to wing the footnote reference, I couldn't automate it either.

As a possible application, consider a time series of velocities measured in high Reynolds number, statistically stationary, turbulent flow.  It is a standard assumption that the power spectrum of such a flow obeys the Kolmogorov $k^{-5/3}$ scaling at intermediate wave numbers. Typically, a suitable log-log plot is used to test this and even to deduce small $\ll {\cal O}(1)$ corrections to the power scaling, caused by fine-scale intermittency. Given the inevitable measurement noise, how clearly is this scaling distinguishable from, say, $k^{-6/3}$ scaling?  The latter is mimicked by the Lorenzian power spectrum whose ACF is exponential, i.e., a first order Markov process.  This is where one could run the $Q$ transform, to fingerprint the time series {\em without prejudice}, at the ``machine learning'' stage, before committing to a stochastic process model.  

%The digital fingerprint obtained here is a minimalist prescription for characterizing a stochastic process. We remark in closing that a more complete ``portrait'' would follow from computing a parallel independent etalon mirroring Figure \ref{fig:gsfig}. 

\begin{figure}
  \centerline{\includegraphics[height=2.8in, ]{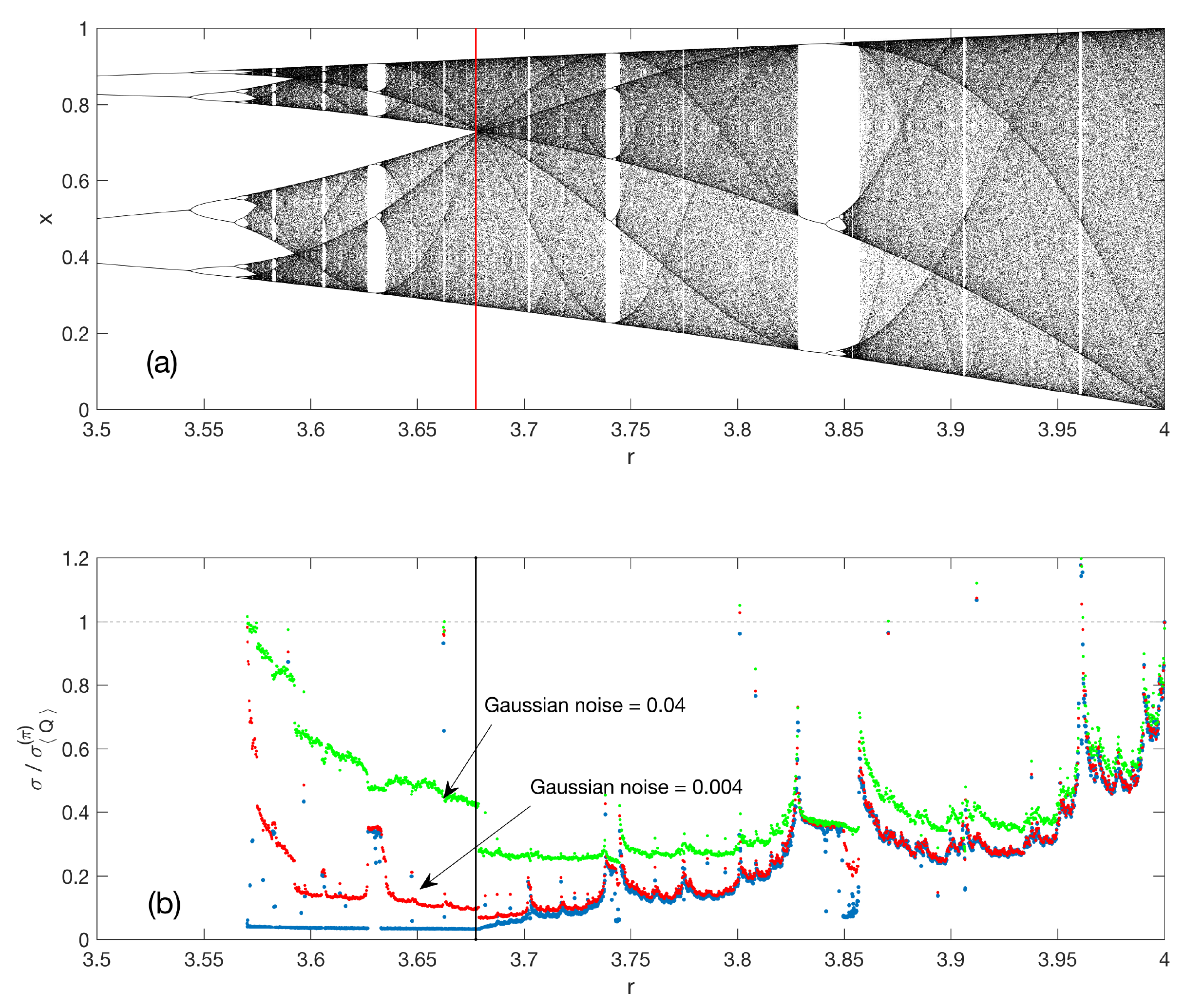}}
  \caption{{\bf Q distinguishes chaos from noise: distribution of $\sigma/\sigma_{\langle Q\rangle}$ for the logistic map:} (a) The logistic map vs.\ $r$. (b) Estimate for the standard deviation for $\langle Q\rangle$, normalized by its value for $\pi$-noise as given in (\ref{eq:meanq_asy}). Each of the lacunae in the map in (a) has its counterpart as an interrupted trace in the curves below. Note the dividing line at $r_c = 3.6875$ marks a boundary between spiked and normal pdfs. (See Fig. \ref{fig:logistic2}.) The lowest trace is that for pure deterministic chaos, the two above show its modification in the presence of additive Gaussian noise with $\sigma = 0.004$ (as in \cite{bandt2002}) and $\sigma =0.04$. Note the separatrix at $r_c$: to the left, the spiked pdfs are more quickly altered by a given stochastic noise level where the normal pdfs of the right responds only slightly.}
  \label{fig:logistic1}
\end{figure}

\begin{figure}
\centerline{\includegraphics[height=3in]{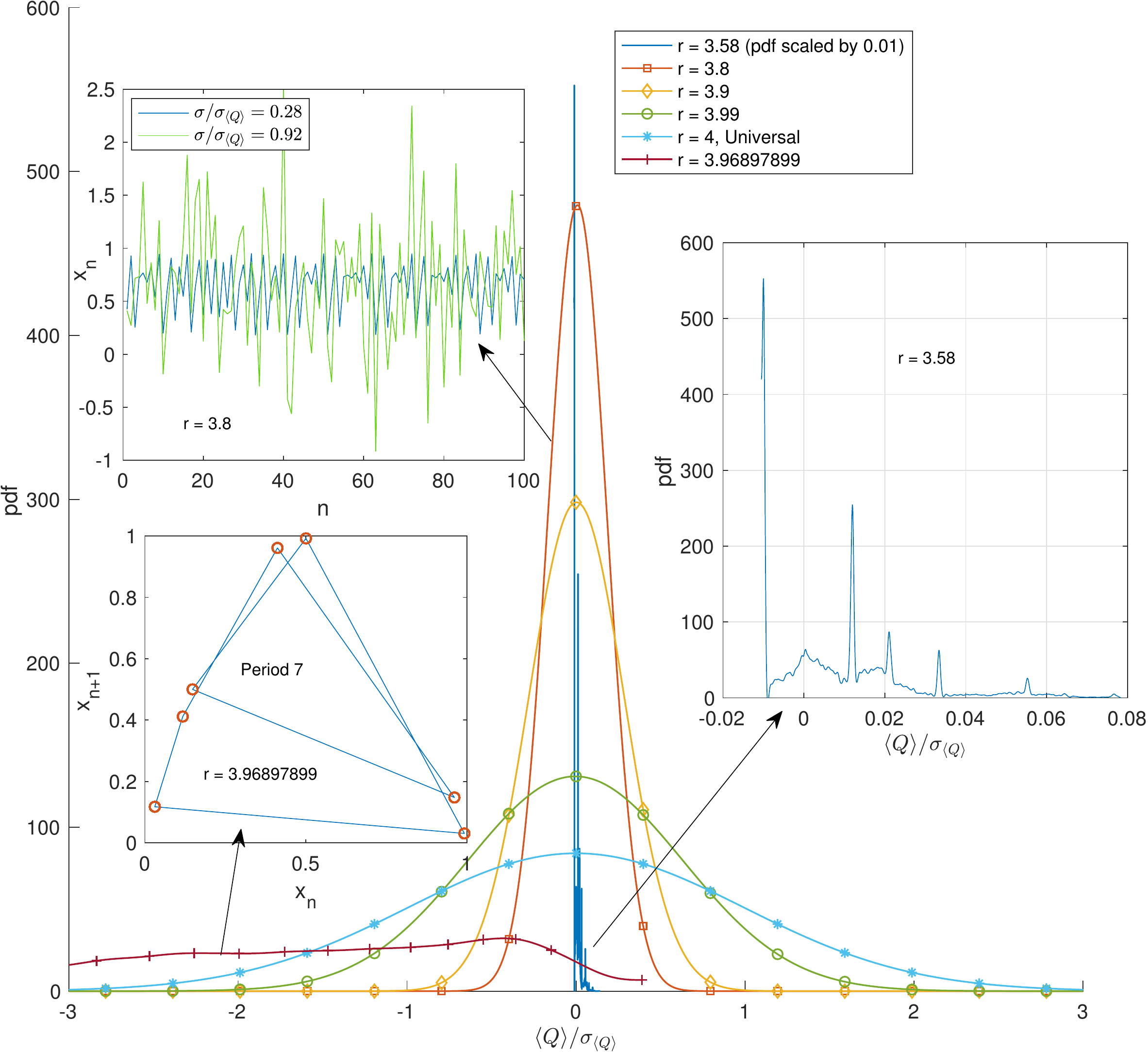}}
  \caption{{\bf $Q$ distinguishes chaos from noise: pdf of $\langle Q \rangle$ vs. $r$:}
  A representative spiked pdf when $r < r_c$ is shown for $r = 3.58$. Note the spiky character. Immediately for $r > r_c$, this gives way to normal distributions as these plotted here for $ r= [ 3.8, 3.9, 3.99, 4]$. The last has a standard deviation exactly matching the prediction from \ref{eq:meanq_asy}, indicating that the output of the logistic map at $r = 4$ exactly matches iid noise statistics. However, the path to this is punctuated by spikes in the $\sigma$ plot at e.g. $r = 3.96897899$ where $\sigma/\sigma_{\langle Q \rangle} = 16 $ (well off scale in this truncated plot). This constitutes a ``normal signal'' in the form of a period seven orbit (lower left). At upper left are two traces: pure deterministic chaos, and the same with added Gaussian noise with $\sigma = 0.5$. Despite intense noise, the chaos imprint is readily discernible.}
 \label{fig:logistic2}
\end{figure}

\section{An illustration from deterministic chaos: the logistic map}\label{sec:logistic}

In many physics applications, ``noise'' is fluctuations in the measurement or observation, e.g., \cite{van1976noise}, while ``signal'' has suggested deterministic components. Chaos produced, for example, by a nonlinear dynamical system is neither.  Following a suggestion by an anonymous reviewer, we digress to test the $Q$ transform on deterministic chaos generated by the famous ``logistic map''
\be
x_{n+1} = r \, x_n \, (1 - x_n) \, ,\label{eq:logistic}
\ee
and show that it compares favorably for detection with the highly regarded ``permutation entropy'' complexity measure of Bandt and Pompe \cite{bandt2002}. 

While earlier we relied upon metrics such as $\langle Q \rangle$ rising {\em above} a threshold value dictated by the desired confidence level as the means for signal detection, with deterministic chaos the tables are turned. A chaotic trajectory is of course, in a loose sense, ``noisy'' but the implication for the pdf of $\langle Q \rangle$ is that it is not noisy enough; it fails to span the gamut of values that would be seen with, say, $\pi$-noise. A general signature of this is that the standard deviation falls below the asymptotic estimate in (\ref{eq:meanq_asy}). When this occurs, we conclude that deterministic chaos is present in the time series, either alone or in concert with stochastic noise.

The bifurcation sequence through which a chaotic map is reached at $r_\infty = 3.569945672$ is discussed in, e.g.\ \cite{cvitanovic2006}. We take a time series from (\ref{eq:logistic}) with $ 128 \times 64$ entries, and reconstitute it in matrix form again with the entries stacked vertically. The pdf for $\langle Q \rangle$ as a function of $r$ is instructive, as seen in Figures \ref{fig:logistic1} and \ref{fig:logistic2}. For $r = 3.58$, immediately above onset at $r_\infty$ the pdf is extremely narrow and multi-peaked. These peaks are vestiges of the principal bifurcation branches at lower $r$. But by $r=3.8$ all such evidence is absent; the pdf is normal with a standard error of $\sigma = 1.45 \times 10^{-4}$. As anticipated, this chaotic data presents with a systematically narrower range of $\langle Q \rangle$ values than found for random noise which, based on (\ref{eq:meanq_asy}), would have  $\sigma_{\langle Q \rangle} = 8.059 \times 10^{-3}$ (the scale factor for the $x$-axis here). But, with increasing $r$, the width grows and, at the end point of $r=4$, the pdf for $\langle Q \rangle$ coincides exactly with the earlier described ``universal distribution'' for noise. This general picture needs to be qualified as suggested by filamentary structure in Fig. \ref{fig:logistic1}. 

The initial transition from a spiked pdf to a normal distribution occurs at $ r_c\approx 3.6875$, as marked by the vertical line in Fig. \ref{fig:logistic1}(a), where upper and lower branch families first meet. As noted by a referee, there is a parallel feature that pairs with this transition in the pdfs for $\langle Q\rangle$; below $r_c$ the pdf for $x_n$ itself is singular, above $r_c$ the pdf, still punctuated with singularities, has full support. Yet another representation of this stochastic  ``phase transition'' is the fingerprint of Section \ref{subsec:fingerprint}, which for the logistic map has a discontinuity at $r = r_c$. 

However there are thereafter discrete departures again from the normal pdf, e.g. those associated with the gaps centered at $r = 3.74$ and $r = 3.84$. There is a large isolated spike at $ r = 3.96897899$ with the indicated anomalously broad pdf, stemming from an orbit of period seven. It achieves a peak of $\sigma/\sigma_{\langle Q \rangle} = 16$, i.e., this represents normal signal detection by $Q$. Similar features punctuate the curve elsewhere. Each feature in Fig. \ref{fig:logistic1}(b) can be linked with associated structure in the logistic map above but the {\em general} pattern, again, consists of Gaussian pdfs of increasing standard deviation to the right. 

Figures \ref{fig:logistic1} and \ref{fig:logistic2} depict the standard deviation for the distribution of $\langle Q \rangle$ with (\ref{eq:meanq_asy}) used as the benchmark for $\pi$-noise. By continuity, near the terminus at $r=4$ and bracketing the spike at $ r = 3.96897899$ must lie two adjacent values of $r$ at which $\sigma/\sigma_{\langle Q \rangle} = 1$. These are not the loci of $\pi$-noise, however, as the coincidence with the value from (\ref{eq:meanq_asy}) is a necessary, but not a sufficient condition. A practical sufficiency condition is that the pdf itself when $\sigma/\sigma_{\langle Q \rangle} = 1$ also pass the Kolmogorov-Smirnov test for normality. This is true only at $r=4$, not elsewhere. 

The red and green traces in Fig. \ref{fig:logistic1}, show the displacement of the curves due to the addition of Gaussian noise of the indicated magnitude. Note the increasingly sharp discontinuity at the $r_c$, with Gaussian (or smooth) pdfs only minimally disrupted by noise while the singular ones exhibit heightened sensitivity. Furthermore, the inset at top left in Fig. \ref{fig:logistic2} shows two traces: the logistic map for $r=3.8$, and the same output with added Gaussian noise of $\sigma = 0.5$. Even for intense noise, a decrease in SNR of $42$ dB relative to the highest noise level used in \cite{bandt2002} -- this combination of signal plus noise remains distinguishable from pure noise as indicated by a standard error $0.92$ that expected from (\ref{eq:meanq_asy}). Indeed as $Q$ is a global method, for any $r$ in the chaotic range, $\sigma/\sigma_{\langle Q \rangle}$ approaches unity only when the stochastic contribution tends to infinity and so for any finite noise and a sufficiently long record, it is always possible to detect a presence of  chaos. Thus, by sensing and transforming rank fluctuations, $Q$ detects subtle aspects of disorder: the distinction between stochastic noise and deterministic chaos. 

\section{A general (nonlinear) regression principle}\label{sec:regress}

With $Q_{rms}$ we have a general purpose, indeed with the extension in Section \ref{sec:extension} to general time series, a universal penalty function as an alternative to least square error. To illustrate this we consider the nonlinear parameter estimation problem of fitting two exponential functions. This is well known as an ill-posed problem for a least squares fit. The classical problem in physics for which this model arises is of course radioactive decay. Though we adopt this setting for its familiarity, multiple exponential fits arise in many other arenas, among them the fitting of transmission functions in radiative transfer and dwell time distributions for ion channels in biophysics. Many special-purpose routines have been written for applications of such multiple exponential fits (e.g., see \cite{wiscombe1977exponential,landowne2013exponential}) and here we consider a representative package, the variable projection method ``varpro'' \cite{oleary2013decay}, and show that $\min Q_{rms}$ outperforms it. But, unlike varpro and other software, e.g. implementation of the Pad\'e-Laplace algorithm \cite{yeramian1987pade}, we change nothing. {\em  We minimize $Q_{rms} $ no differently than we would in fitting a noisy quadratic curve.}  There are no parameters to tune, no weights.\footnote{One downside in compared with high order regressions of {\em linear} least square problems: the latter is solved by simple matrix inversion. Use of $Q_{rms}$ gives a nonlinear minimization problem, though a robust one for all cases we have explored.}
%Occasionally, we encountered that minimizing the least square error can be at odds with reducing $Q_{rms}$ and in higher order can actually {\em increase} $Q_{rms}$ to several times above the 95${}^{\rm th}$ percentile noise limit.}

\begin{figure*}
 \centerline{\includegraphics[height=2.75in]{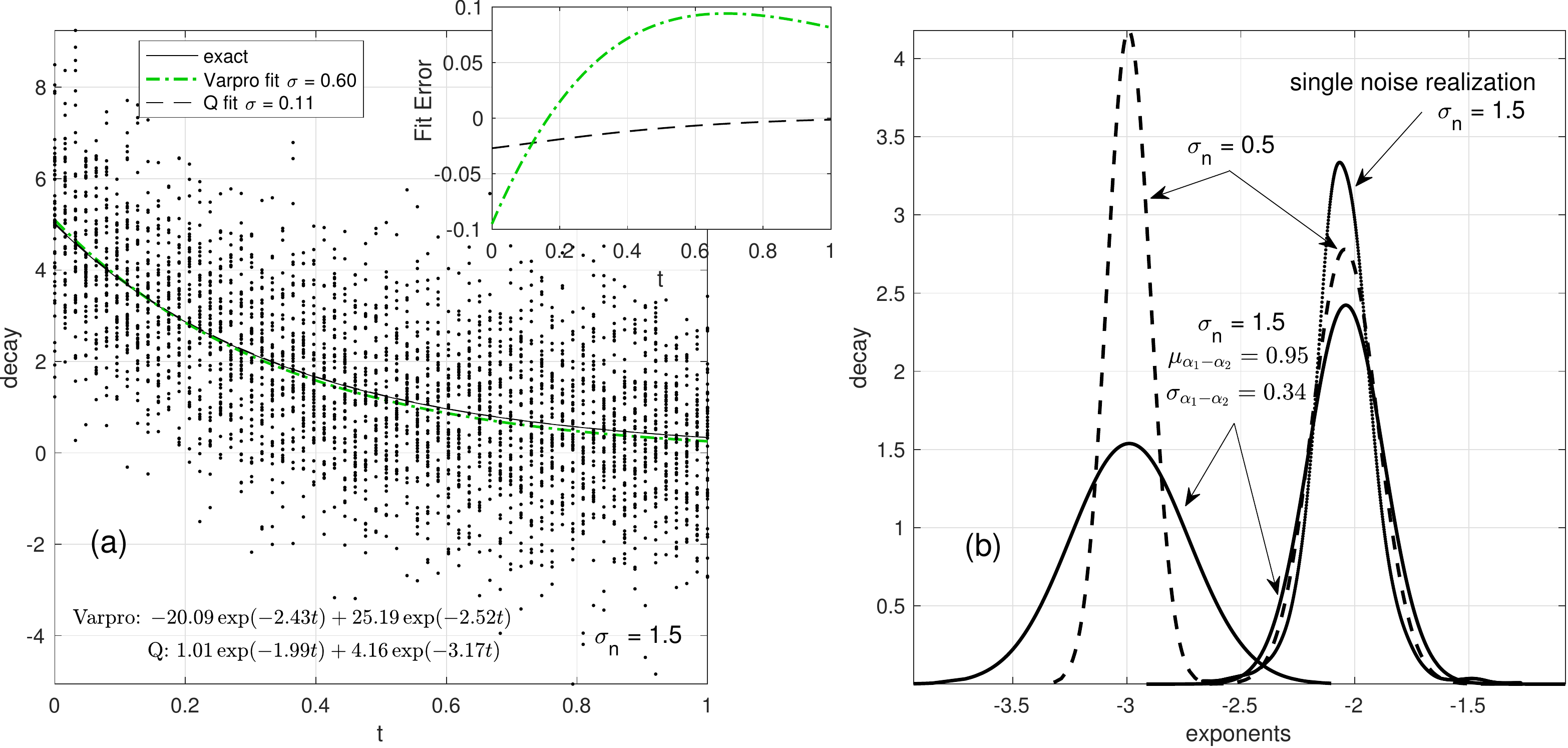}}
\caption{{\bf Nonlinear Regression for the Two-Species Radioactive Decay:} (a) Data is the sum of two decaying exponentials buried in Gaussian white noise with $\sigma = 3/2$. Also shown are $Q$ and Varpro fits and the true signal. The indicated standard errors for $Q$ are relative to the exact answer and on this basis the $Q$ regression is five times more accurate. But the deeper problem is revealed by the detailed form of the regression for this realization (bottom), where the varpro result settles for two nearly equal exponents and large coefficients of opposing signs while $Q$ matches all four parameters well. (b) The pdfs for each $Q$ fit exponent from 2500 realizations, solid black for noise with $\sigma = 3/2$, dashed for $\sigma = 1/2$. The dotted curve shows the pdf for repeated independent minimizations {\em of a single realization}, a consequence of the imbricated surface seen in the next figure.}\label{fig:radio1}
\end{figure*}

\begin{figure*}
 \centerline{\includegraphics[height=2.75in]{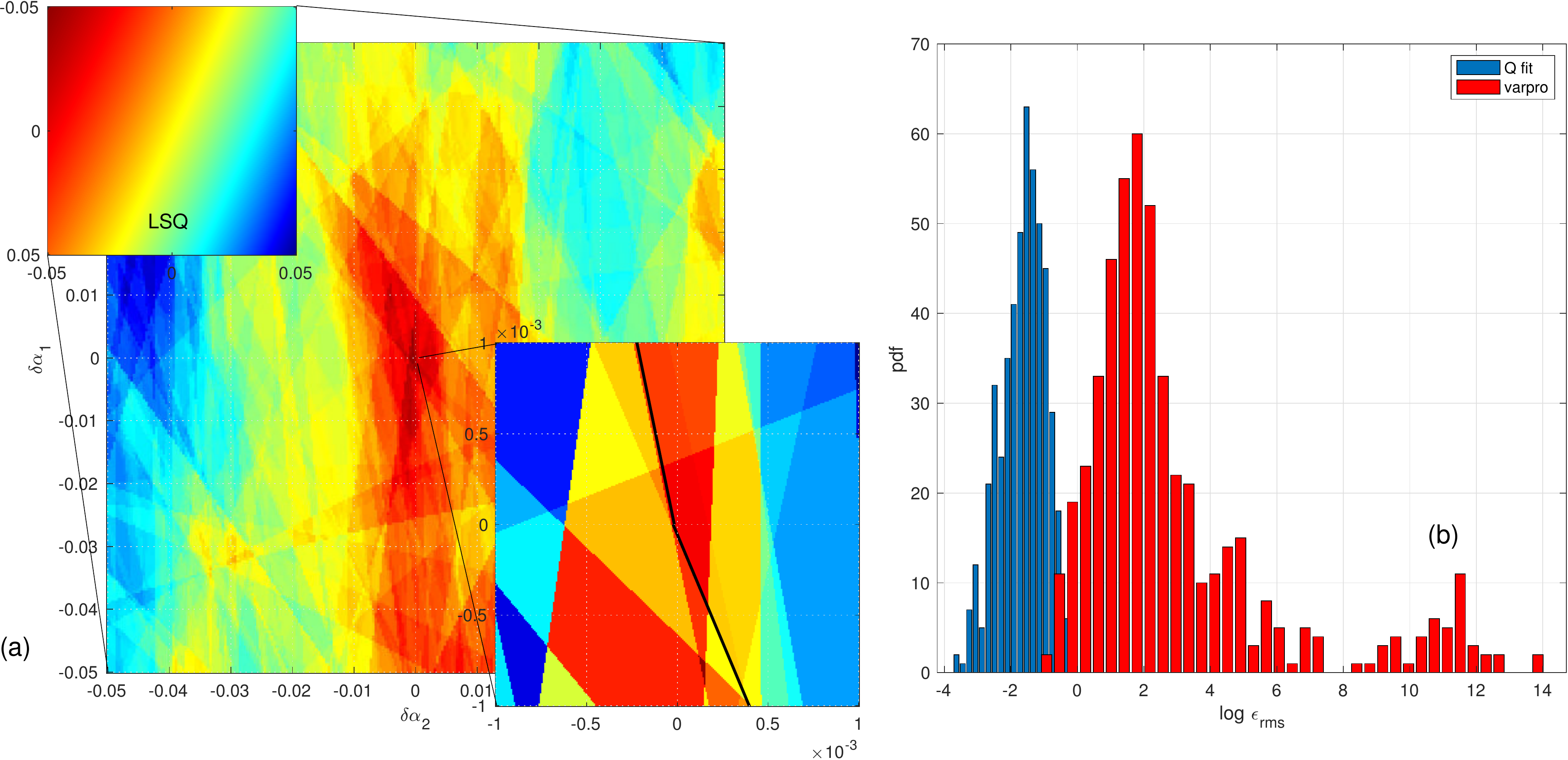}}
\caption{{\bf $Q_{rms}$ Optimization of the Decay Exponents:} (a) Main image shows $1/Q_{rms}$ for the single realization above (maxima - red - are more readily visible than minima - blue), plotted  in the $(\alpha_1, \alpha_2)$ plane, and centered about the exponent pair listed in Fig \ref{fig:radio1}(a). The discreteness of rank creates this mosaic; a palimpsest of wedges. The Nelder-Mead simplex method is suitable here but in a given search gets trapped by one of many nearly identical local minima. Zooming in around the origin $\times 50$ shows the simplex vertex where the algorithm converged but, almost directly below this, a tiny simplex that (slightly) exceeds this maximum. (Color for the inset is rescaled.) To the upper left is a plot for the entire region of the reciprocal of the conventional LS error, with no local maximum at all. Rather, for this $(c_1,c_2)$ pair, the global LS maximum is at $\alpha_1 = -2.699$ and $\alpha_2 = -3.055$, significantly worse and far out in the tail of the dotted pdf of Fig. \ref{fig:radio1}.  (b) Regarding the unknowns as a four-vector ${\bf v} =[c_1, c_2, \alpha_1, \alpha_2]$, with the exact solution denoted by ${\bf v}_0$, we compute $ \epsilon_{rms} \equiv | {\bf v} - {\bf v}_0 | $ for varpro and $Q$ fits in a Monte Carlo simulation of 500 trials. Only about 5\% of the former lie to the left of the worst single $Q$ fit.}\label{fig:radio2}
\end{figure*}

Consider a signal of the form 
\be
C(t) = c_1 \, \exp (  \alpha_1 \, t) + c_2 \, \exp (  \alpha_2 \, t) \label{eq:radio}
\ee
with $c_1 = 1, c_2 = 4$ and $\alpha_1 = -2, \alpha_2 = -3$ and the observations consist of 50 repeated ``measurements'' taken at 64 evenly spaced points on the interval $t = [0,1]$. For so short an interval and given relatively close exponents, even the noise-free fitting problem can be challenging. Here we complicate the situation greatly by the addition of Gaussian noise with $\sigma = 3/2$. As seen in Fig. \ref{fig:radio1}(a), the raw data show only a general exponential decay; there is no immediate indication of two species. Varpro requires seed values for the exponent pair. Conservatively, (to give varpro maximum advantage), {\em in all cases} we seed with the exact values.  Values for the coefficients and exponents based on $Q$ proceed very much like the earlier process of detrending. One takes initial values for these, substitutes them into (\ref{eq:radio}) and subtracts the resulting values of $C(t)$ from each of the realizations in the data matrix. The $Q_{rms}$ of the residual is computed and then minimized by varying the vector of unknown parameters. We used the Nelder-Mead Matlab routine fminsearch for that minimization.

For the result of the single realization in Fig. \ref{fig:radio1}(a), the $Q$ regression also has been seeded with exact values. While the $Q$ regression does fit the exact result better, the main point about exponential fits is that the varpro result is a fairly good fit as well. But, where the $Q$ fit yields reasonably accurate coefficients and exponents, the varpro coefficients are wildly in error, of opposite signs, with a meaningless negative value. 

In Fig. \ref{fig:radio1}(b) we see the Gaussian pdfs for the standard error of each $Q$-determined exponent both for $\sigma = 3/2$ and also $\sigma = 1/2$. Each of these is a projection from a four-dimensional pdf. One side effect of that projection is an apparent modest overlap of the two exponents around the value of $-2.5$. If one steps back to the two-dimensional pdf projection that obtains in the $(\alpha_1, \alpha_2)$ plane, near coincidence of values becomes a negligible fraction. The sample mean value of $\alpha_1 - \alpha_2$ is $ 0.95$ with a standard deviation of $0.34$, so near equality occurs only at the 3 sigma level. A final revealing (non-normal) pdf is that plotted with dots for $\alpha_1$. Here are the values obtained with repeated invocations of the Nelder-Mead routine using random perturbations of the starting seeds about their exact values. One does {\em not} obtain a single well-defined minimum, rather there are countless, nearly equal, local minima clustered in a small region leading to a pdf with sample mean of $\mu = -2.049$. From a slice through the 4-D volume of $Q_{rms}(c_1, c_2, \alpha_1, \alpha_2)$, taking in particular the $(\alpha_1, \alpha_2)$ plane, one sees in Fig. \ref{fig:radio2}(a) a finely structured field with numerous overlapping wedge-shaped regions. (Maxima are more easily discerned with this color map so $1/Q_{rms}$ is plotted.) Optimization with this simplex structure needs an appropriate routine, and the Nelder-Mead algorithm proves well suited. In the magnified view (inset at lower right), one can see that, while the algorithm has settled on a simplex vertex that is a local maximum, it missed the better tiny simplex almost directly beneath. These issues are local; {\em all} the exponent values for $\alpha_1$ found with the randomly perturbed initial seeds are reasonably accurate, moreover their standard error foreshadows the Monte Carlo simulation with independent realizations of noise. Note that optimization routines customarily allow for user set tolerances. One of these is the function tolerance; how small a change of $Q_{rms}$ is realizable. Given that $Q_{rms}$ derives from rank, this is a discrete value. The smallest possible change is found by perturbing the center of the $P$ matrix with the following $ 2 \times 2$ matrix:
\[
\begin{pmatrix} 
+1 & -1 \\
-1 & +1
\end{pmatrix}
\]
This manifestly preserves the row and column sum identities and consists of a rank exchange of one in two adjacent entries. For the model problem here that leads to $ \Delta Q_{rms} = 4.78 \times 10^{-8}$. Finally, in Fig. \ref{fig:radio2}(b) we compare the varpro and $Q$ results for the Monte Carlo simulation. The results of the former are so poor that one cannot compare exponent to exponent and coefficient to coefficient. Instead, we adopt a simple gross measure. We let  ${\bf v}_0 = [c_1, c_2, \alpha_1, \alpha_2]$ and use ${\bf v}$ to denote the vector with components given by their numerically determined values.  We then compute $\epsilon_{rms} \equiv | {\bf v} - {\bf v}_0 |$ as a measure of the error. The dynamic range is so large that we plot the distribution of the log this quantity. About 5\% of the varpro results are slightly better than the {\em single worst} $Q$ result, and can be sensibly associated with the expected values of coefficients and exponents. About 25\% of the remainder consist of solutions similar to that listed in Fig. \ref{fig:radio1}(a); two nearly equal exponents and coefficients that satisfy  $c_2 \approx 5 - c_1$, with $c_1 < 0$. For the remaining 75\%, one exponent is about $ -2.5$ and the second is much larger in magnitude. The latter are evenly split between large positive values with coefficients of order $10^{-6}$ and large negative values with coefficients of order one. All of these results, except the initial 5\%, amount to the same conclusion about the data; that there is only a single exponent present. 

We have assumed it is known that: (1) exponential decay is the correct model, and (2) two species are present. One could assume a state of complete ignorance, but we think it fair at least to assume exponential decay is understood to be the relevant model. But, one may well not know {\it a priori} the number of species. There is, as a reviewer noted, then no basis on which to prefer the Varpro or the $Q$ result. As they use different metrics, all one can say is that each has minimized what was asked of it. But there is a difference. Varpro, or any other software that relies upon least square error for the penalty function, is incapable of stably fitting more than a single exponent for data with this level of noise. The $Q$ fit, by contrast, offers a single exponent fit of $5.0445\, \exp( -2.6552\, t)$, and a stable two exponent fit. However, the values of $Q_{rms}$ are essentially identical -- $0.021664$ (one species) and $0.0216591$ (two species) --  so one cannot on that basis prefer one solution over the other. Other evidence is required.

No more is needed for the practical application of $Q_{rms}$ in a multitude of other problems. One simply replaces a routine that computes the least square error of a trial regression with one that returns $Q_{rms}$ for the trial. Error bounds are desirable in any application, but one cannot give a universal characterization for these, even for least square applications. For a linear trend, one can obtain a general form for the standard error of the slope and this is done for the $Q$ fit in Section \ref{sec:extension}. 

%We turn next to a decidedly more heuristic part and present initial evidence for a conjecture inspired by the group PCA results. 

\section{Signal extraction from noisy data without {\it a priori} knowledge of signal shape}\label{sec:dqdt}

The opening example in Section \ref{sec:deriveQ} established a surprising result: that rank data, lacking all magnitude information, can nonetheless predict linear trends in noisy data in excellent agreement with slopes found from the traditional (unweighted) least squares. In this section we argue for a far stronger result: the same rank information yields an assumption-free estimate for a general nonlinear signal with relative amplitude information intact. 

Inspired by the close correspondence of undulations in the $\psi_k$ modes and oscillations in the companion $\delta \mu_k$, we propose that, up to a linear rescaling, the underlying signal is well approximated  by $-d \overline{Q}(t)/dt $, where the overbar denotes the mean over rank in $Q$ (i.e., horizontal mean).\footnote{Note a tentative parallel result for signal extraction of nonstationary variance, namely $\delta\sigma(y) = \int^y \, dy'' \, \int_{-1}^1 \, dx'' \, x'' \, q(x'',y'')$.} The need of linear rescaling arises because rank is invariant under $f(t) \to  \alpha f(t) + \beta$.

Evidently it is the differential impact of systematic rank arising from signal juxtaposed against random rank scrambling by the noise that allows for the signal magnitude recovery. But this depends upon a finite signal-to-noise ratio (SNR); the limit of a perfect input signal is singular and the recovered signal in that limit is (counterintuitively) less accurate.

\begin{figure}
\centerline{\includegraphics[height=2.75in]{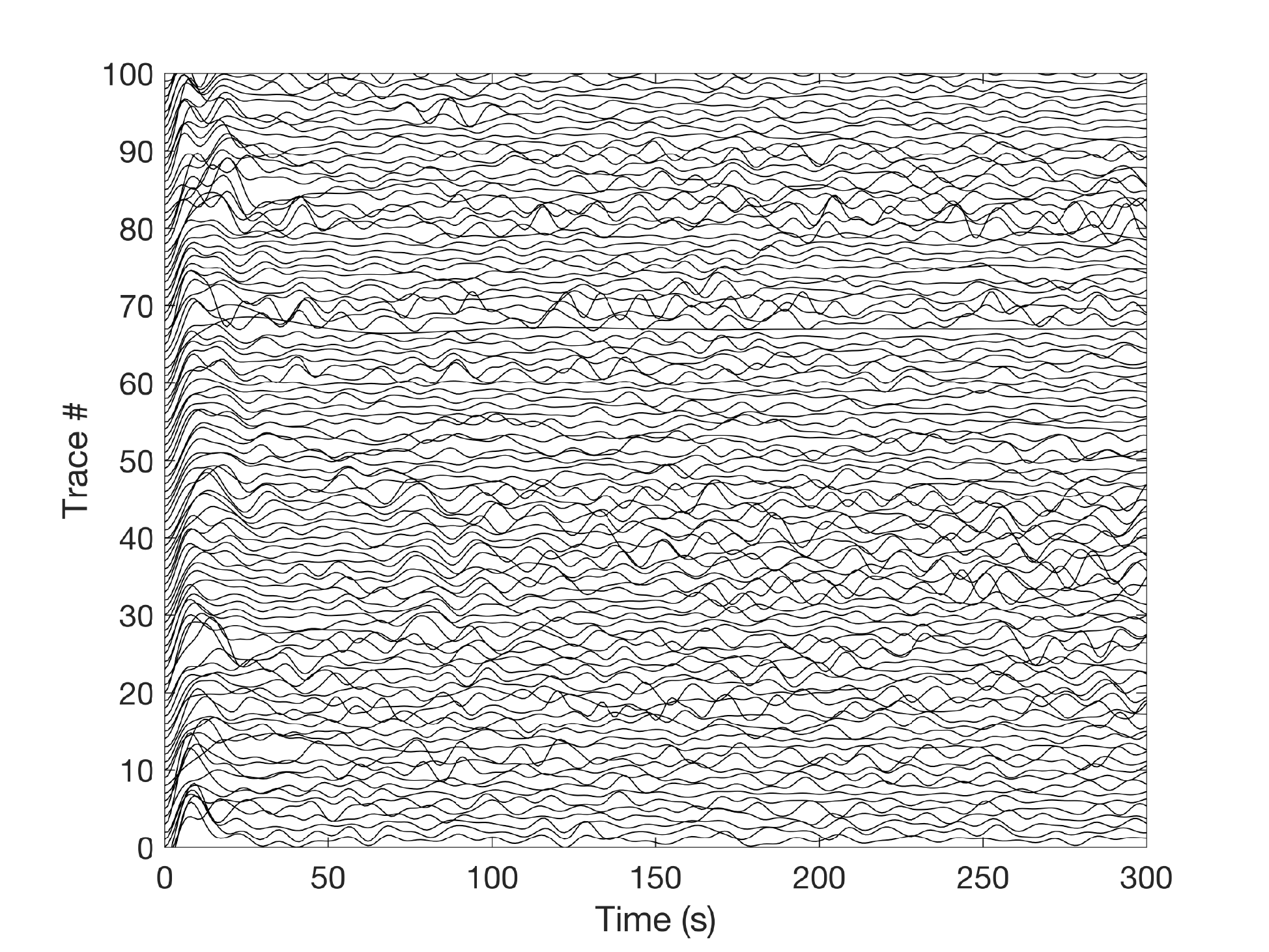}}
\caption{{\bf Raw seismic S-wave amplitudes:} The first 100 of 1192 traces from the USArray between $96$ and $97$ degree epicentral distance.}\label{fig:dqdt}
\end{figure}

\begin{figure}
\centerline{\includegraphics[height=2.4in]{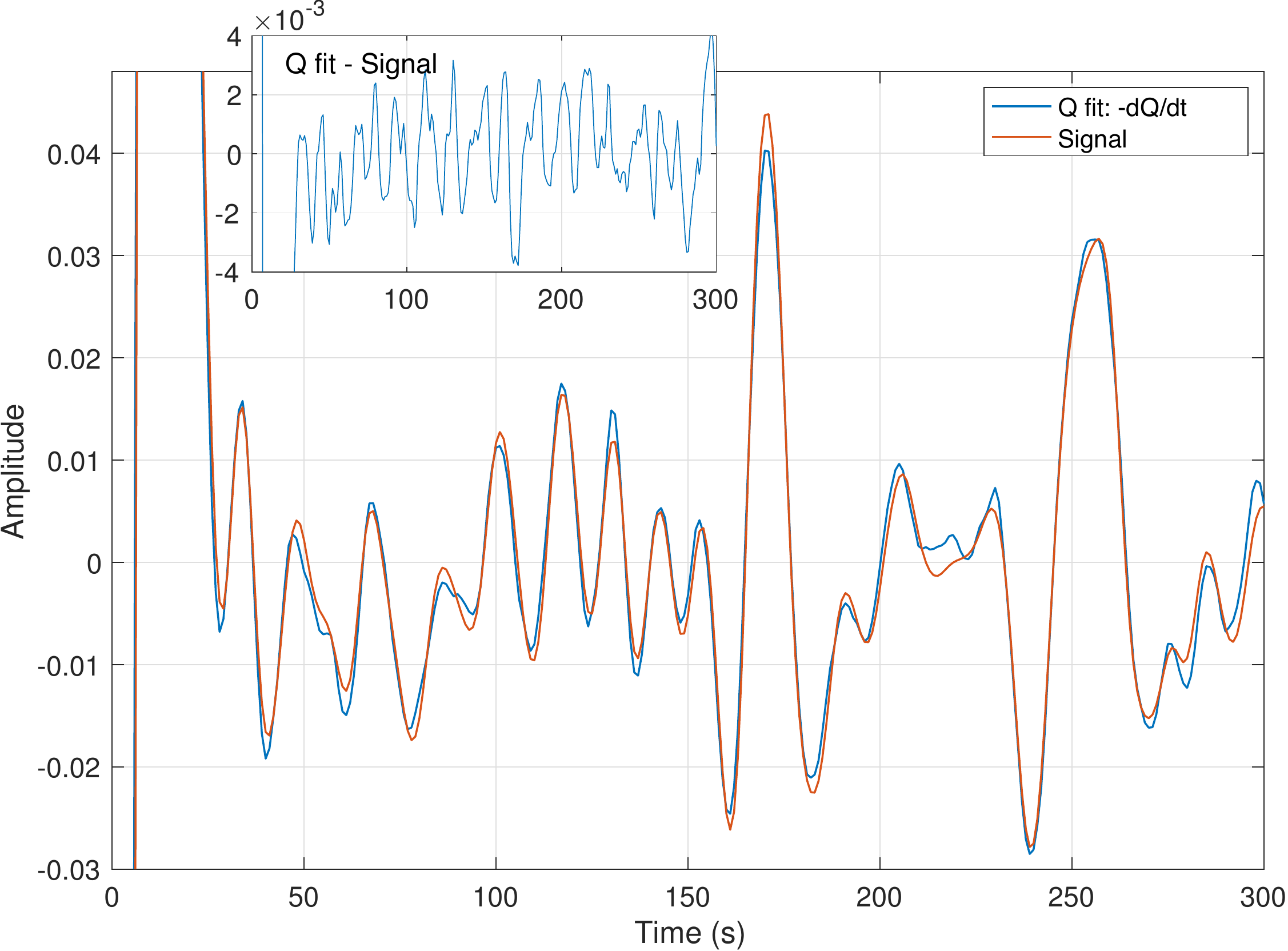}}
\caption{{\bf Signal extraction from  $-d \overline{Q}(t)/dt$:} 
The conventional method (red curve) is to average the noisy individual 1192 traces illustrated in Figure \ref{fig:dqdt}. The resulting S-wave peaks at 171 and 255 seconds are linked to mantle discontinuities at 440 and 660 km. The assumption-free, {\em ordinal} signal extraction (blue curve) from $-d\overline Q/dt$ matches remarkably well, particularly the phase.}
\label{fig:dqdt2}
\end{figure}

We are indebted to Professor Peter Shearer of UCSD for the raw data from a forthcoming publication\cite{shearer2019}, a sample of which is shown in Figure \ref{fig:dqdt}. The full set consists of $n_t = 1192$ S-wave reflections from the 410- and 660-km mantle discontinuities between 96 and 97 degree epicentral distance. Here $n_T = 301$, the data are uniformly spaced at $\Delta t = 1$s and the authors of \cite{shearer2019} use the mean over 1192 traces as the signal proxy. In Figure \ref{fig:dqdt2} we compare that signal with the result from $-d\overline Q/dt$ with the difference between the two shown in the inset. The two spikes centered at 171 and 255 seconds correspond to reflections at the above noted 410- and 660-km mantle discontinuities respectively. As noted in \cite{shearer2019}, the oscillations are part of a signal rather than noise as these do not decrease as $n_t^{-1/2}$, ($n_t = 1192$, \# of traces), and the $Q$-approach confirms this independently, just as it distinguished chaos from noise in Section \ref{sec:logistic}. 

For this comparison, the free linear rescaling of $-d\overline Q/dt$ was chosen to match the arithmetic mean most closely. (In a general application without a reference signal, the multiplicative scale $\alpha$ can be set by minimizing $Q_{rms}(\alpha)$. Here that dependence is fairly weak with a shallow minimum that gives a similar result.)  The new result from $-d\overline Q/dt$ shows excellent fidelity with the benchmark: the {\em phase} of all the oscillations is spot on, the differences are confined to small changes in peak amplitudes. 

As with the initial result for Lugano, where we found a slope from annulling $\langle Q\rangle$ of
$2.4875 \degree$C over 65 years, nearly identical to the standard LS fit of  $2.5165 \degree$C per 65 years, so too here we obtain a result nearly identical to one previously found by more conventional methods. The initial message is, we reiterate, that this agreement is achieved based solely on rank information. Just as for Lugano where we expanded the reach of the $Q$ transform to nonlinear parameter estimation and (in the next section) to data fitting in the presence of heavy tail noise, with results in each case unmatched by conventional methods, so too here we anticipate that signal extraction with $-d\overline Q/dt$ offers comparable opportunities. 

\section{$Q$ performs well in heavy-tailed noise\label{sec:heavytail}}

So far mostly Gaussian white noise has been used but here we examine distributions with heavy tails, where outliers are ubiquitous. In the least square family these are often handled with the bisquare method, which excludes outliers adaptively by assigning them zero weight. However, for distributions with infinite mean and/or variance a more powerful approach is needed. We are grateful to an anonymous reviewer for suggesting a comparison with the Theil-Sen estimator, used exclusively to determine linear trends \cite{balkema2018risks}. Its potential limitation is computation time for large data sets. For example {\em each} trial of $365 \times 64$ data pairs for Table \ref{table:heavy} required 20 seconds of CPU time on a 2.5 GHz Intel Core i7 laptop. The full implementation requires fitting slopes to all possible pairs of points, which takes ${\cal O}(N^2)$ operations. Several theoretical papers have proposed ${\cal O}(N \, \log N)$ implementations but no public code, so far as we know, is available, although CPU time may not be a practical concern for small to medium scale applications.

\begin{table}
\begin{ruledtabular}
\begin{tabular*}{\hsize}{@{\extracolsep\fill}lccc@{}}
%\topline
Distribution & $Q$ $\sigma$ & LS $\sigma$ & Theil-Sen $\sigma$ \\
\colrule
\ Uniform  $ (-1,1)$ & 0.012 & 0.007 &  --- \\
\ Gaussian $(\sigma = 1, \mu = 0)$ & 0.023 & 0.023 & --- \\
\ Cauchy   $(\sigma = 1, \mu = 0)$ & 0.042 & 0.038 &  ---  \\
\ Pareto   $(x_m = 2/3, \alpha = 2/3)$ & 0.022 & 0.068 & 0.023\\
\ GEV  $(\xi =2, \sigma =1, \mu =0)$ & 0.016 & 27.84 &  0.018\\ 
%\botline
\end{tabular*}
\end{ruledtabular}
\caption{Comparison of the standard error for a linear trend with unit rise in 64 years in: $Q$, bisquare-weighted robust least square estimator and, for the two most challenging cases, the Theil-Sen algorithm ($10^3$ trials).  The GEV result for bisquare least square has numerous severe outliers for slope estimates. Similarly, for the slightly modified form $ x\, \exp(-x)$ with GEV noise, bisquare least squares yields $ 2.82 \pm  2.81$, $Q$ gives $1.003\pm  0.047$, and the Theil-Sen estimator is inapplicable.\label{table:heavy}}
\end{table}

In Table \ref{table:heavy}, the first two cases pose no problem, even for unweighted least squares, though we quote the bisquare result for consistency. The Cauchy distribution is the first point where the bisquare adaptive approach becomes critical; unweighted least squares is not useful. But then even the bisquare method begins to lag in performance for the Pareto distribution and finally is unusable for the generalized extreme value distribution. Both of the latter distributions have infinite mean and variance. By contrast, for Pareto and GEV, Theil-Sen performs admirably as expected. But so does detrending by simply setting $\langle Q \rangle = 0$, which is as earlier noted already a practical ${\cal O}(N \log N)$ algorithm. While the GEV distribution may seem a far-fetched choice, in fact it arises in applications such as analysis of hydrometeorological data for maximum precipitation events \cite{adluoni2005rain}. 

One can generalize this problem slightly to the multilinear form, $ c_1\,  x_1 + c_2\,  x_2$.\footnote{Or, for that matter, $ f_1(x;c_1) + f_2(x;c2)$.} To take a practical example, set $c_1 = 3$ and $c_2 = -2$ for a $65 \times 90$ grid. For the case of Pareto noise from the $Q$ fit we obtain $c_1 = 2.9994 \pm 0.0425$ and $c_2 = -2.000 \pm 0.0429$. The bisquare algorithm reports $ c_1 = 2.9995 \pm 0.0850$ and $-2.000 \pm 0.0772$ so, as in Table \ref{table:heavy}, it is beginning to fray. In contrast, there is no parallel procedure for the Theil-Sen test. There is an unpublished manuscript by Dang et al.\ \cite{dang2008theil} for the multilinear case, but it remains an unrealized routine for general application. So, for the multilinear case with GEV noise, neither method offers a result to compare with the $Q$ fit of $c_1 = 2.9997 \pm 0.0323$ and $ c_2 = -2.0000 \pm 0.0326$.  

Note also about the general multilinear problem that the $Q$ regression is unusual compared to one's experience based on the least square formulation. We  obtain $c_1$ and $c_2$ {\em individually} by setting $\langle Q \rangle = 0$ twice; once for the data matrix in each orientation. This generalizes to a multilinear form in {\em any number} of variables with the slight modification that one has first to appropriately permute, and then to reshape, the matrix for each of the coefficients to be determined. This decomposition is possible because $Q$ is invariant to a constant offset.

\section{On the general application of $Q$ \label{sec:extension}}

Although the $Q$ transform was developed for regularly spaced data such as that in Fig. \ref{fig:1abcd}(a), it is flexible in application and here we touch upon the possibilities. For example, uniform spacing of the temperature data by day and year could be replaced by recording daily low temperature when first attained, i.e., by the continuous astronomical Julian date, including the hour, minute, and second. Then the abscissae are irregularly spaced.   

\begin{figure}
\centerline{\includegraphics[height=3.75in]{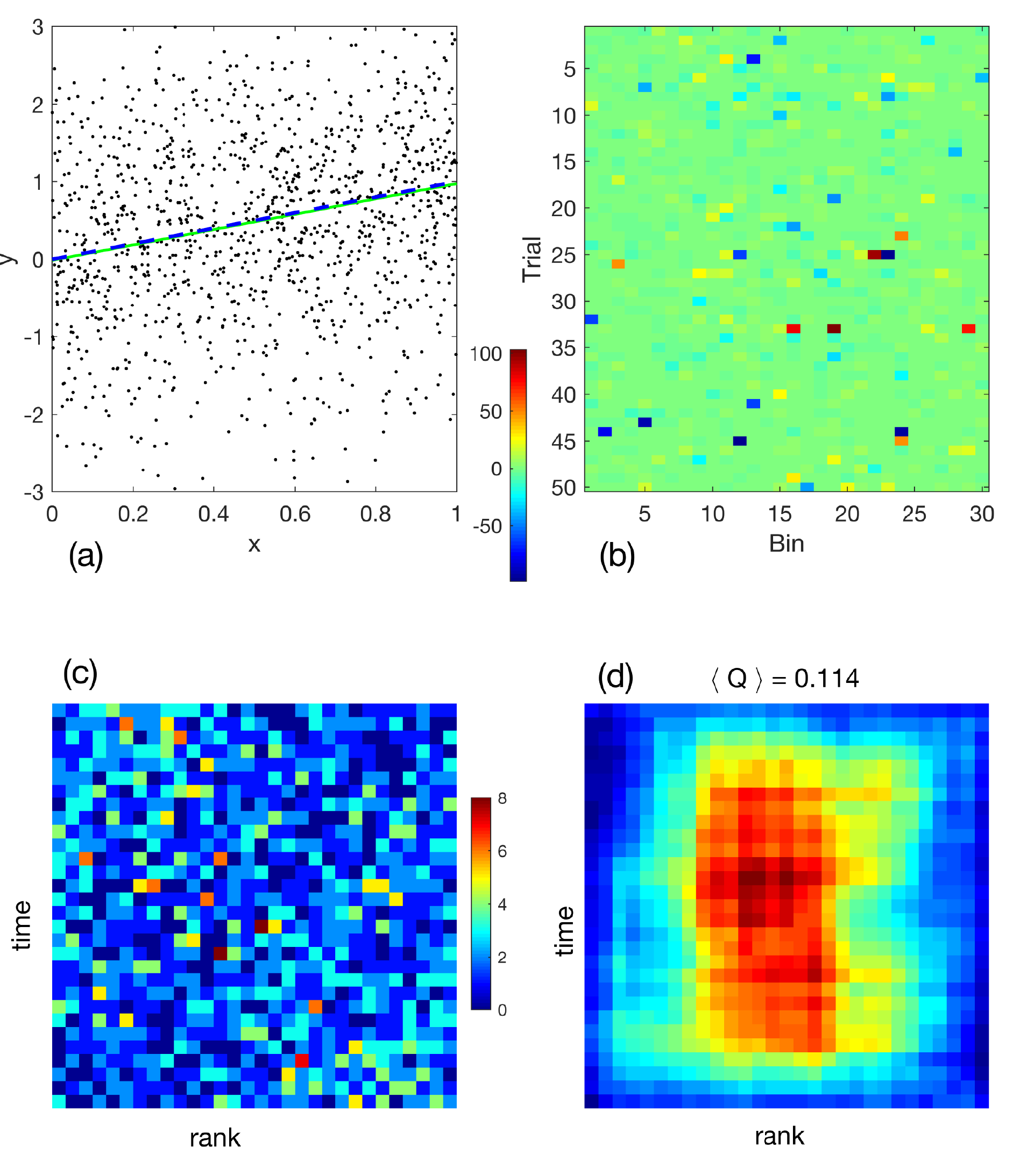}}
\caption{{\bf Application of $Q$ to general data fitting:} 
(a) A sample of 1500 data points ($x$ values) drawn from a uniform distribution, The $y$ values: linear trend plus noise from a Cauchy distribution. The vertical range is truncated to show the local fit but the actual data range over $[-100,100]$. (b) While randomly space sample points seem far from the initial application of linear regression with $Q$, by suitably grouping these points we obtain the equivalent of Fig. \ref{fig:1abcd}(a); (c): $P$ matrix; (d) the $Q$ matrix. $Q$ and, more so, $P$ clearly show the imposed linear trend. The $Q$ fit trend (solid line), supplemented by a local algorithm to estimate the constant term, compares well with the exact trend (dashed line). }\label{fig:general}
\end{figure}

A model data set is plotted in Fig. \ref{fig:general}(a) consisting of 1500 $(x_k,y_k)$ pairs. The $x_k$ coordinates are generated from a uniform distribution on the interval $[0,1]$. The $y_k$ values are given by
\be
y_k = x_k + n_k \label{eq:lin}
\ee
where $n_k$ are noise values from a Cauchy distribution with mean $\mu = 0$ and scale $\sigma = 1$. 

We need a data matrix from which to compute $P$ and $Q$. For this purpose we subdivide the $x_k$ into $M$ bins each with $N$ points, with $M\times N = 1500$. We choose comparable $M=50$ and $N=30$. From the asymptotic formula at (\ref{eq:meanq_asy}), to leading order there is no difference if these are reversed, and the numerics confirm it. But (\ref{eq:meanq_asy}) is asymptotic and one cannot approach either extreme (e.g. 1500 bins and 1 experiment, or vice-versa) without a breakdown in the formalism. $M$ should be chosen with the Nyquist frequency in mind whenever information about the expected signal is available. 

Now we partition the $x$ data into the 30 bins; the first bin containing the 50 smallest values of $x$, $\ldots$ , up to the last bin with the 50 largest. Each $x_k$ is then paired with its corresponding $y_k$, so bin \#1 now contains 50 $(x,y)$ pairs, and so on. Finally, we assemble the 50 trials. For experiment 1,  take element 1 ($y_1$),  from bin \#1, element 1 ($y_{51}$) from bin \#2, etc.  Up to the 50${}^{th}$ experiment: take the last remaining element from bin \#1 ($y_{50}$) etc.

The resulting data matrix is shown in Fig. \ref{fig:general}(b) along with a color scale to show the wide range of values associated with this noise distribution. From one row to the next, the $x$ coordinate in a given column is no longer constant (previously the calendar year) but, as each experiment is independent, nothing hinges upon that constancy; each row still represents a linear trend that we attempt here to estimate as a function of the horizontal coordinate. For the purposes of computing $P$ and $Q$, that horizontal coordinate is the (integer) bin number, while if a specific functional relation (trend) is to be tested, we appeal to the specific $x_k$ for that row and column. 

Proceeding to enumeration of $P$ we obtain the $30 \times 30$ matrix illustrated in Fig. \ref{fig:general}(c). Although noisy, the $P$ matrix has a bias, with upper left and lower right overpopulated, indicating a positive trend. The $Q$ matrix at right confirms this. Here $\langle Q \rangle = 0.114$ and this can be compared to the noise benchmark at (\ref{eq:meanq_asy}) on the assumption that the latter does indeed hold for all distributions. For the present $M$ and $N$ one obtains $\sigma_{\langle Q \rangle} = 0.0196$ and $\langle Q \rangle$ here is well above the noise level; there is a signal. Moreover $Q_{rms} = 0.1395$ and the ratio $| \langle Q \rangle / Q_{rms} | = 0.8193 $ is very close to the ratio of $0.8341$ for a pure linear signal with no noise,  evaluated at $30$ points, indicating that the signal, based on $Q_{rms}$, has nearly the maximum trend possible based on $\langle Q \rangle$.

We annul $\langle Q \rangle$ exactly as before taking care that, when the trend is computed, matrix entries must be computed individually since columns of the raw data matrix are no longer at fixed $x$. The result is a slope estimate of $0.9869$ hence an error of $0.0130$. The result is plotted as a solid line in Fig. \ref{fig:general}, the exact result is dashed. 

The generalized $Q$ fits are insensitive to a constant offset and one has to find another tool for that purpose. For heavy-tailed distributions a local method of matching the estimated sample is preferable. On mild assumptions about noise statistics, one expects a fitted line to lie in the dense ``middle'' of a cloud of sample points. A simple method to estimate that middle is first to subtract the $Q$ fit and then to count the sample points lying within a sliding window, fixing the intercept as the midpoint of the window location where the convolution peaks. We used the first member of a Slepian sequence \cite{slepian1978prolate, thomson1982spectrum} for that window successfully, yielding the intercept for the solid line in Fig. \ref{fig:general}(a).

This exercise repeated 1500 times gives an estimate for the mean slope error of $0.0060$, consistent with a limiting value of zero, that is, an unbiased estimator. It also gives an estimate for the standard deviation of the slope of $0.171$. Note for this latter that  
\[
\frac{\sqrt{30 \times 50}}{\sqrt{64 \times 365}} \, 0.171 = 0.043\, ,
\] 
consistent with the value of $0.042$ reported in Table \ref{table:heavy}. 

\subsection*{Data matrix considerations \& error estimates\label{error}}

We have constructed raw data matrices reflecting a variety of origins of noise and signal. The simplest circumstance is $\delta$-correlated noise, with successive rows representing $n_t$ repeated trials, and convergence to the underlying signal scaling as $n_t^{-1/2}$. Typically $n_T$ would be determined by the expected signal duration or period. In most instances, either $\langle Q\rangle$ or $Q_{rms}$ would serve as the metric, and their general asymptotic expansions are as indicated in (\ref{eq:meanq_asy}) and (\ref{eq:qrms_asy}). Beyond this, from (\ref{eq:pform0}) and (\ref{eq:pform}) we have the analytic foundation to demonstrate that for such noise in the absence of signal, the ensemble average $P$ is constant and hence $Q$ vanishes. 

Often, however, the noise is stationary but correlated. As shown in Appendix \ref{sec:exactP}, the ensemble average of $P$ is no longer constant and hence the mean $Q$ is nonzero. But, since the induced
$P$ has $D_4$ symmetry, the ensemble mean of $\langle Q \rangle$ is still zero.
One must still compute the modified standard deviation to set appropriate thresholds for signal detection. While the mean of  (\ref{eq:meanq_asy}) is altered by the $D_4$ corner effects on $P$, this diagnostic has anyway a nonzero mean in all cases, so Monte Carlo computations will automatically adjust for this correction. Still, in some cases it may make sense to use an altered $\tilde Q_{rms}$ based on $\tilde P$, with its $D_4$ projection removed.
 
The raw data matrix in the introductory climate example manifests another phenomenon. Here the data was wrapped vertically in the matrix, that is, the end of column 1 then continues on at the top of column 2, and so on. Here not only do we have the vertical correlation whose effects were considered in Section \ref{sec:stats} but, from the theoretical perspective of Appendix \ref{sec:exactP}, one would need to model as well the cross-correlation {\em as between columns}. All the more must we rely upon numerical evidence. As first noted in Section \ref{sec:deriveQ}, the principal effect of (positive) {\em vertical} correlation is a reduction in the effective value of $n_t$. Extensive numerical simulation further indicates that, in spite of cross-column correlation, the ensemble mean $P$ remains constant in the absence of signal provided the columns are long enough, relative to the correlation length, so that row elements are uncorrelated. 

A variant of this issue arises if one seeks to extract not the long term, but the seasonal signal. Then the original data matrix for Lugano is turned 90 degrees so it is the end of one {\em row} which is correlated with the start of the next. Again the ensemble mean of $P$ for pure noise is {\em not constant}, and so the ensemble mean of $Q$ is not zero. While this again represents a potential bias, numerical results suggest the ensemble  mean $P$ still has $D_4$ symmetry and hence does not affect estimated trends in the mean, only estimated trends in variance.\footnote{For chaos a nonvanishing ensemble mean of $Q$ is general even with rows of independent trials. The deterministic nature of such processes invalidates a result like (\ref{eq:pform0}).}

With this preamble, we turn to the practically important question of error analysis in the simplest case of iid noise. As remarked previously, when looking for estimated slope error, one has to restore the link between the rank-order space of $Q$ and the dimensional space of the raw data. Now the underlying pdf of white noise -- Gaussian, Cauchy, etc.  affects the slope error.  If the noise in (\ref{eq:lin}) is replaced by Gaussian noise with standard deviation $\sigma$ then for $K= M\times N$ total sample points, the large $K$ limit of the standard deviation of the (unit) slope for an unweighted least squares fit is
\be
 \sqrt{\frac{12}{K}} \, \sigma \approx \frac{3.46\, \sigma}{\sqrt{K}}
 \label{eq:OLS}
\ee
For the present procedure we can appeal to the leading term of (\ref{eq:meanq_asy}), which must then be divided by the ensemble average of $d\langle Q \rangle / d\alpha$ evaluated at $\alpha = 0$ to calibrate the change in the mean value of $Q$ when perturbed by a signal $\alpha \, x$. It is through this factor that the connection between the particular noise distribution and signal manifests itself, accounting for the variation of the $Q$ entries in Table \ref{table:heavy}. In principle this derivative can be computed analytically by taking the mean value of the $Q$ transform of the Fr\'echet derivative of (\ref{eq:pform}). Short of that, direct {\em numerical} evaluation of that Fr\'echet derivative for $M = 15$ leads to $d\langle Q \rangle / d\alpha = 0.2422/\sigma$. This can be compared in a test of consistency to a numerical fit from Monte Carlo simulations for varying $M$ of
\be
\frac{1}{\sigma} \, \left [ 0.2117 + 0.1618\, \exp(-0.1087\, M)\right ]\, ,\label{eq:MC}
\ee
which gives $0.2434/\sigma$ at $M=15$. Taking the large $M$ limit of (\ref{eq:MC}) and the leading term in (\ref{eq:meanq_asy}) then gives the standard deviation of the slope estimate as
\be
\frac{0.7131 \, \sigma}{0.2117 \, \sqrt{K}} = \frac{3.37 \, \sigma}{\sqrt{K}}
\label{eq:QLS}
\ee
hence LS and $Q$ fits of slope are, for this Gaussian case, essentially identical. As noted in the introduction, LS is the maximum likelihood estimator for this case hence one cannot improve upon (\ref{eq:OLS}). That the constant in (\ref{eq:QLS}) is slightly smaller is not however a contradiction. Rather, it reflects a compounding of errors from two delicate estimations for asymptotic constants, namely (\ref{eq:MC}) and (\ref{eq:meanq_asy}).

Strictly speaking, (\ref{eq:meanq_asy}) only applies for a discrete set of $n_T$ abscissae, not to the larger generalized set of $K$ points here. But for the above estimate we need only a leading order result and for that it suffices to use (\ref{eq:meanq_asy}) with the abscissae chosen as the column-by-column means. 

The procedure above extends readily to fitting an unknown signal by minimizing $Q_{rms}$ using an expansion in a basis set of the user's choosing. One can extend the binning here to higher dimension and then parallel the development of Section \ref{sec:regress}. Lastly, one can  pursue the second half of the $Q$ formalism, with $ - d\overline{Q}/dt$, but this is beyond the scope of this paper.

%Stationary processes that depart from $\pi$ noise have, in contrast, zero means for these modal pdfs with, however, one significant exception. For $\delta$-correlated white noise, the ensemble-averaged $P$, necessarily of $D_4$ symmetry, is a constant matrix with entries of $n_t/n_T$, as can be proven from (\ref{eq:pform0}) for example. With any departure from perfect $\delta$-correlation, however, (\ref{eq:pform0}) and (\ref{eq:pform}) no longer apply. While no longer constant, the ensemble-averaged $P$ is still of $D_4$ symmetry. Of the precursor modes of $P$ that obtain from inversion of the $\psi_k$ of Figure \ref{fig:gsfig}, exactly {\em two} have the required $D_4$ symmetry: those from $\psi_6$ and $ 1/\sqrt{2} \, ( \psi_{19} + \psi_{20}) $, modes of $R_2$ symmetry for $Q$. These $Q$ modes hence play two distinct roles for stationary stochastic processes: modification of width of each distribution reflects increased (or decreased) transient fluctuations in standard deviation relative to $\pi$ noise (e.g. the quadratic profile $\delta\sigma_6(x)$), while modification of their {\em means} away from zero reflects departure of that stochastic process from pure $\delta$-correlation. Such autocorrelation is better characterized by other measures, as discussed below. 

\section*{Concluding Remarks}

The ordinal nature of the $Q$-transform introduced in this paper gives it great versatility, extending to time series with different units and to imaging and rendering it robust with respect to gaps in data.  The algorithm is simple, objective, and fast.  It performs well in various types of noise, including heavy-tailed. 

The unknown signal, whether deterministic or random, is defined by the departure from the ``equality of ranks'', that is, uniformity (constancy) of the ensemble-averaged rank population matrix $P$.  At a single realization level, the departure is from the `salt-and-pepper'' $P$ (Poisson process). The logic is reminiscent of the first law of thermodynamics: when introducing internal energy, one does not yet know what ``heat'' is, but understands its {\em absence} through heat insulation.  Remarkably, this ``not noise'' definition readily distinguishes deterministic chaos from noise as illustrated on the data produced by the logistic map even in the presence of significant white noise. That same fingerprint which, for some, serves to detect signal can, for others, serve to revise the noise threshold against which some {\em other} signal is then judged.

For parameter estimation, even the linear model is on first glance surprising as annulling $\langle Q \rangle$ accurately recovers the slope despite having no magnitude information; only ranks, perturbed by noise. The keys to understanding emerge in the group PCA analysis of Section \ref{sec:svd} and the error analysis of Section \ref{sec:extension}. With the extension to  minimization of $Q_{rms}$, one has then a general alternative for least square error as a penalty function. Success with the canonical nonlinear parameter estimation problem of two-species radioactive decay and excellent performance for heavy-tail noise without need of empirical weights are harbingers of promise for future applications. 

Using the $Q$ transform for nonparametric signal extraction without prior information on signal shape in a blind and distribution-independent manner is documented with the seismological data of Figure \ref{fig:dqdt}. The required, heuristic, form $-d\overline{Q}/dt$ is quickly and easily computed. While more theoretical development is needed, exploratory applications will be of great interest.

We hope that the reader will try these ideas as the Matlab code is supplied in supplemental material.

\acknowledgments
This work was supported by the NSF grant AGS-1639868. We thank anonymous reviewers for suggesting comparison with the Theil-Sen algorithm, and the logistic map and Ornstein-Uhlenbeck process as test cases.

\appendix

\begin{figure} % first figure, Appendix A
\centerline{\includegraphics[height=3.6in]{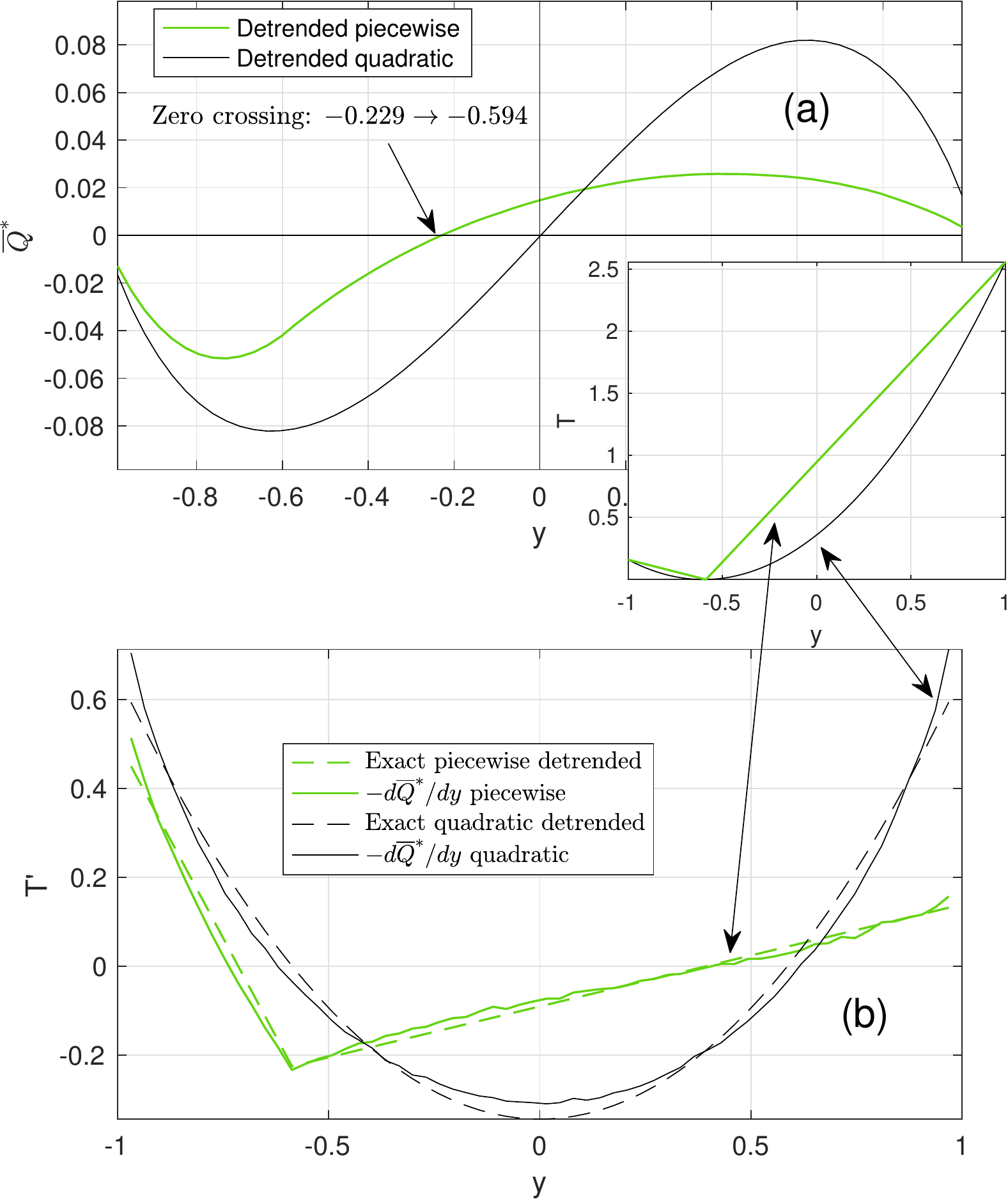}}
\caption{{\bf Quadratic vs.\ piecewise linear profiles and break detection:} Two profiles are shown in the inset: the quadratic, $(y+3/5)^2$ and a piecewise linear profile matching the endpoint values and zero minimum at $y = -3/5$. (a) ensemble means $\overline{Q}^*$ computed for iid normal noise with $\sigma = 2$. The antisymmetric cubic profile is consistent with the detrended quadratic, which is centered at the origin, while the zero crossing of $\overline{Q}^*$ for the detrended piecewise case maps accurately using (\ref{eq:bimodal}) to $-3/5$. (b) $-d\overline{Q}^*/dy$, when linearly rescaled, works well for {\sl both} detrended profiles. If a bimodal pattern of $Q$ emerges after detrending with a zero crossing of $\overline{Q}$ significantly displaced from the middle, this is, likely, a discontinuity in the time series at the indicated node (as inferred in Fig. 11), caused by e.g., changes in thermometry or station location. In such cases relying on $-d\overline{Q}/dt$ from a single realization is less robust than use of (\ref{eq:bimodal}), with no differentiation to fix the node. (A significant cubic component in the profile can {\em also} displace the zero crossing but this typically shows in the annual mean.)}\label{fig:riddle}
\end{figure}

\section{\label{sec:breaks}Detecting breaks in a time series}

Another, more quantitative, prediction follows from (\ref{eq:qcont}) by noting that once a linear trend is removed from a data set, the residual $Q$ is often a double-lobed horizontal structure of alternating sign. This is the signature of a correction to the temperature profile with alternate periods of cooling and warming, but no net trend. Similar bimodal patterns arise in $Q$ after detrending either a quadratic temperature profile or a piecewise linear version, but with significant differences as shown in Fig. \ref{fig:riddle}. 

\begin{figure} % second figure, Appendix A
\centerline{\includegraphics[height=3in]{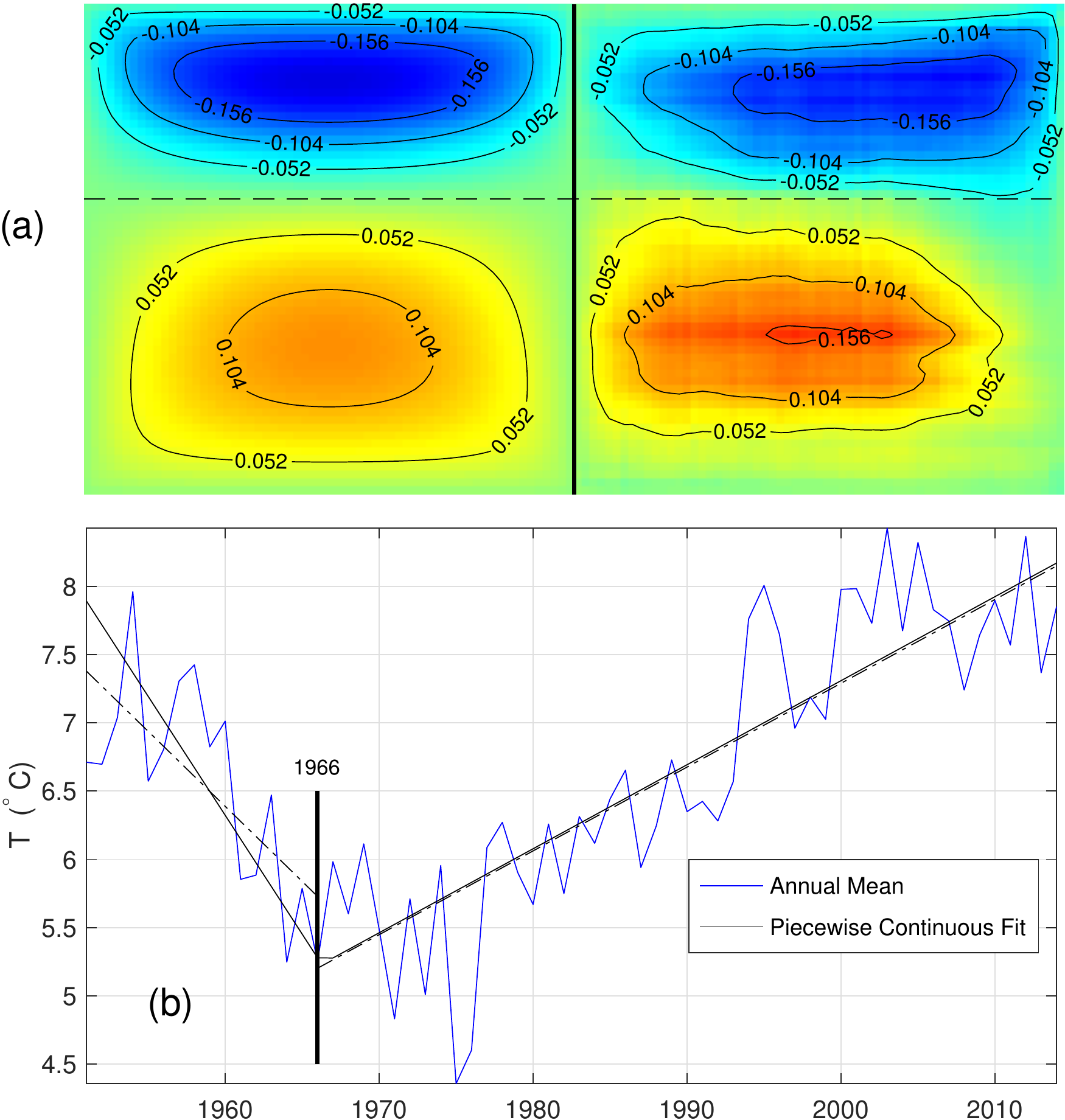}}
\caption{{\bf Inference from $Q$ of a break in slope:} (a) The slight left-right asymmetry in the pattern at top right (USW00023050 Albuquerque Int'l Apt, NM) is a harbinger of Mode 2 in Figure \ref{fig:gsfig}, with a downward trend in $\sigma$ of about $0.4 \degree$C. (b) $Q$ and LS detrending of the left hand segment alone give a different slope than results from this piecewise continuous correction, thereby suggesting a breakpoint. }\label{fig:qzero}
\end{figure}

A simple algebraic representation of $P$ for the piecewise case may be taken as a trendless, zero mean, piecewise linear profile in $y$ with a node at $y_n$, multiplied by $x$. Application of (\ref{eq:qsimp}) then yields
\be\label{eq:qbimodal}
\begin{split}
q(x,y)=&\frac{1-x^2}{2\, (y_n+1)\, (y_n+2)\, (y_n-1)^2\, (1 -y^2\, x^2)} \times
\\
&\left [ 2\, (y-y_n)^2\, (H(y_n-y)-H(y-y_n))\right .\\ 
&\left . +y^2\, y_n^3-3\, y^2\, y_n+2\, y\, y_n^2-y_n^3+2\, y-y_n \right] \, .
\end{split}
\ee
A typical pattern for (\ref{eq:qbimodal}) is seen at top left in Fig. \ref{fig:qzero}. To the right is the residual $Q$ after removing the linear trend for station USW00023050 (Albuquerque Int'l Apt, NM). The dashed line is a zero contour of $Q$ on the left, chosen to coincide with the zero of the horizontal mean of the $Q$ at the right. 

A self-consistent way to achieve a breakpoint is simultaneously to detrend each of $Q_{L}$, $Q_{R}$, and $Q$ as well as possible, while requiring that the trend used for the third be the net slope of left and right segments joined as a continuous function. This imposes a jump condition. Each of the three mean values for $Q$ is first weighted by $\sqrt{n_T}$ to put them on an equal footing and their sum of squares then minimized. The dash-dot line in Fig. \ref{fig:qzero} shows the optimal result. The continuity constraint yields a jump of $0.53 \degree$C for raw temperatures to the left of the break. We note a similar isolated nearby empirical breakpoint in 1961, with an estimated bias of $0.5 \degree$C to the left, for the  monthly mean data for this station in the Berkeley Earth series (\#173069). While the coincidence of the bias estimates is striking, the onset date here has to be refined since the deduction based on (\ref{eq:bimodal}) assumes a piecewise continuous profile. A simple trial confirms that a jump moves $y_n$ earlier.

From (\ref{eq:qbimodal}) follows the exact general result that
\be
y_0 = \frac{y_n}{2 + | y_n |}\label{eq:bimodal}
\ee
where $y_0$ is the zero line of $\overline{Q}$. It follows then that all zeros of a bimodal $Q$ must always lie in the middle one-third of the domain. For a timespan of 64 years, that amounts to the middle 21. From a sample of 79 GHCN stations with unbroken temperature records, 40 exhibited an evident bimodal pattern after detrending. In {\em all} cases, the zero of the horizontal average of the residual $Q$ observes this constraint, moreover the relation above then furnishes an objective location for the break point in a piecewise temperature correction, leaving only its amplitude to be determined. For the case illustrated, the zero of $Q$ is at the beginning of 1976, in accord with the middle-third rule, and the indicated node is hence early in 1966.

\section{\label{sec:asympt}More on analytic results}

\subsection{An expression for the ensemble mean of $P$\label{sec:exactP}}

To gain a deeper understanding of $Q$ transform properties (e.g., the signal extraction conjecture $-d\overline{Q}^*/dt$ or normality of the distribution for $\langle Q \rangle$, discussed in later sections), we note here an exact general result for the ensemble mean of $P$ in the case of uncorrelated iid variables with a secular component $T_k$ ($k$ a year index from $1$ to $K$), namely:

\be 
\begin{split}
&P_{n,k}(T) = n_t \,  \int_{a}^b \, dt \,  \mbox{pdf}(t - T_n) \, 
\sum_{j=1}^{{}_{K-1}C_{k-1}} \\
&\qquad\prod_{n=1}^{k-1} \mbox{cdf}( t - T_{{}_js_n} )
\prod_{m=1}^{K-k} \left ( 1 - \mbox{cdf}(t - T_{{}_j\tilde s_m})\, \right )
\end{split}\label{eq:pform0}
\ee
where $\mbox{pdf}$ and $\mbox{cdf}$ are the governing probability density and cumulative  distribution functions on the interval $[a,b]$ with appropriate parameters as needed. Here ${}_{K-1}C_{k-1}$ is the binomial coefficient, $s$ is a matrix whose rows contain all possible choices of  $k-1$ elements from the set $\{ 1,2,\ldots ,K\}_{\sout{n}}$ and
\[
\{{}_j\tilde s_m\} \equiv \{ 1,2,\ldots ,K\}_{\sout{n}} - \{{}_js_n\} \, .
\]
For the useful particular case of Gaussian random component with standard deviation $1/\sqrt{2 \beta}$, the $(n,k)$ element of $P$ is given by
\begin{equation}
\begin{split}
P_{n,k}(\beta | T) &= n_t\, \sqrt{\frac{\beta}{\pi}} \, \left (\frac{1}{2} \right )^{K-1} \, \int_{-\infty}^\infty \, dt \,  \exp(-\beta \, (t - T_n)^2) \,  \\
&\sum_{j=1}^{{}_{K-1}C_{k-1}} \prod_{n=1}^{k-1} (1 + \mbox{erf}(\sqrt{\beta} ( t - T_{{}_js_n})))\\
&\qquad \prod_{m=1}^{K-k} ( 1 - \mbox{erf}(\sqrt{\beta}(t - T_{{}_j\tilde s_m}))) \, .
\end{split}\label{eq:pform}
\end{equation}
A test of this prediction for a linear $T$ against the mean from Monte Carlo trials with $N$ realizations gives a residual with rms error that decays as expected, like $N^{-1/2}$.  The Fr\'echet derivative of these forms proves a central ingredient in error bounds for linear regression. We return to this point in Section \ref{sec:extension}, where the $Q$ transform is broadened to general time series. It would be useful to generalize the equilibrium form (\ref{eq:pform0}) to correlated noise but even the uncorrelated Gaussian case in (\ref{eq:pform}) is difficult, e.g., proving that $P_{n,k}(\beta | 0) = n_t/K$ in the absence of any signal is a complex task of integration and combinatorial identities. Moreover as it stands, owing to the factorial growth of terms, (\ref{eq:pform0}) and (\ref{eq:pform}) are computationally feasible only out to $K \approx 16$, smaller than needed in practice. An asymptotic expansion is needed.

\subsection{Effects of correlation: end effects on $P$}

Even for a stationary random process, correlation introduces a surprise: the ensemble average of $P$ is no longer constant. One can see the origin of this by considering a time series of exactly three entries, $[x,y,z]$. If these are iid with the standard normal distribution (zero mean, unit variance), then the joint pdf for this set is given by
\be
p_1(x,y,z) =
\frac{1}{4\,\sqrt{2\, \pi^3}}\, \exp\left (-(x^2+y^2+z^2)/2\right )\, .
\ee
From the symmetry of this form alone it follows that the probability for each variable being the lowest rank is $1/3$. Numerical experiments suggest this conclusion holds for any iid distribution, a result which may be strengthened by appeal to the argument in \cite{foster1954distribution},  which notes that reshuffling records destroys any rank correlation in a time series. 

We introduce correlation in the simplest possible fashion. Let $x = x_1 + x_2$, $y = x_2 + x_3$, and $z = x_3+x_4$ where $x_{1,2,3,4}$ are iid normal variables as above. Now $x$ is correlated with $y$, and $y$ with $z$, but $x$ and $y$ are uncorrelated. Now the joint pdf is
\begin{equation}
\begin{split}
p_2(x,y,z) &= \frac{1}{4\,\sqrt{2\,\pi^3}} \, 
\exp\left ( -y^2/2 - 3\,(x^2+z^2)/8 \right. \\
&\left. +y\,(x+z)/2 -x\, z/4\right )
\end{split}
\end{equation}
where
\begin{equation}
\begin{split}
  \int_{-\infty}^\infty dx\, p_2(x,y,z) &=  \frac{1}{2\pi\sqrt{3}}\,
   \exp\left ( ( y\, z -y^2 -z^2) /3 \right )\\
  \int_{-\infty}^\infty dz \, p_2(x,y,z) &= \frac{1}{2\pi\sqrt{3}}\,
   \exp\left ( ( x\, y -x^2- y^2) /3 \right )\\
  \int_{-\infty}^\infty dy \, p_2(x,y,z) &= \frac{1}{4\pi}\,
   \exp\left ( - ( x^2+ z^2) /4 \right )
\end{split}
\end{equation}
and one sees in the last three forms the correlation relations stated above. And now the computation for lowest rank yields
\begin{equation}
\begin{split}
\int_{y=x}^\infty &\, \int_{z=x}^\infty \, \int_{x=-\infty}^\infty  p_2(x,y,z) \, dx\, dy\, \, dz \\
&= \int_{y=z}^\infty \, \int_{x=z}^\infty \, \int_{z=-\infty}^\infty  p_2(x,y,z) \, dx\, dy\, \, dz = 3/8
\end{split}
\end{equation}
and
\be
\int_{z=y}^\infty \, \int_{x=y}^\infty \, \int_{y=-\infty}^\infty \, p_2(x,y,z) \, dx\, dy\, \, dz = 1/4\, ,
\ee
with an overshoot at the ends and a low in the middle. The symmetry breaking here is that $y$ is correlated with two neighbors, $x$ and $z$ with only one. For an extended row of this same construction, that symmetry breaking remains confined to the ends. A similar result obtains for the highest rank. 

In consequence, for correlated stationary noise, all four corner regions of $P$ are affected, while the interior approaches constancy. For progressively larger $P$, the fractional area affected tends to zero and so also then the induced ensemble average of $Q$. This effect manifests as a pure $D_4$ contribution to $P$ and pure $R_2$ for $Q$ and so leaves trends completely unaffected. In cases where a variance signal is sought, one could first simulate the noise in a Monte Carlo computation, obtain the ensemble average $P$, and then remove its zero-mean projection on all realizations with variance signal present. 

\begin{figure} % figure for Appendix B
 \centerline{\includegraphics[height=2.75in]{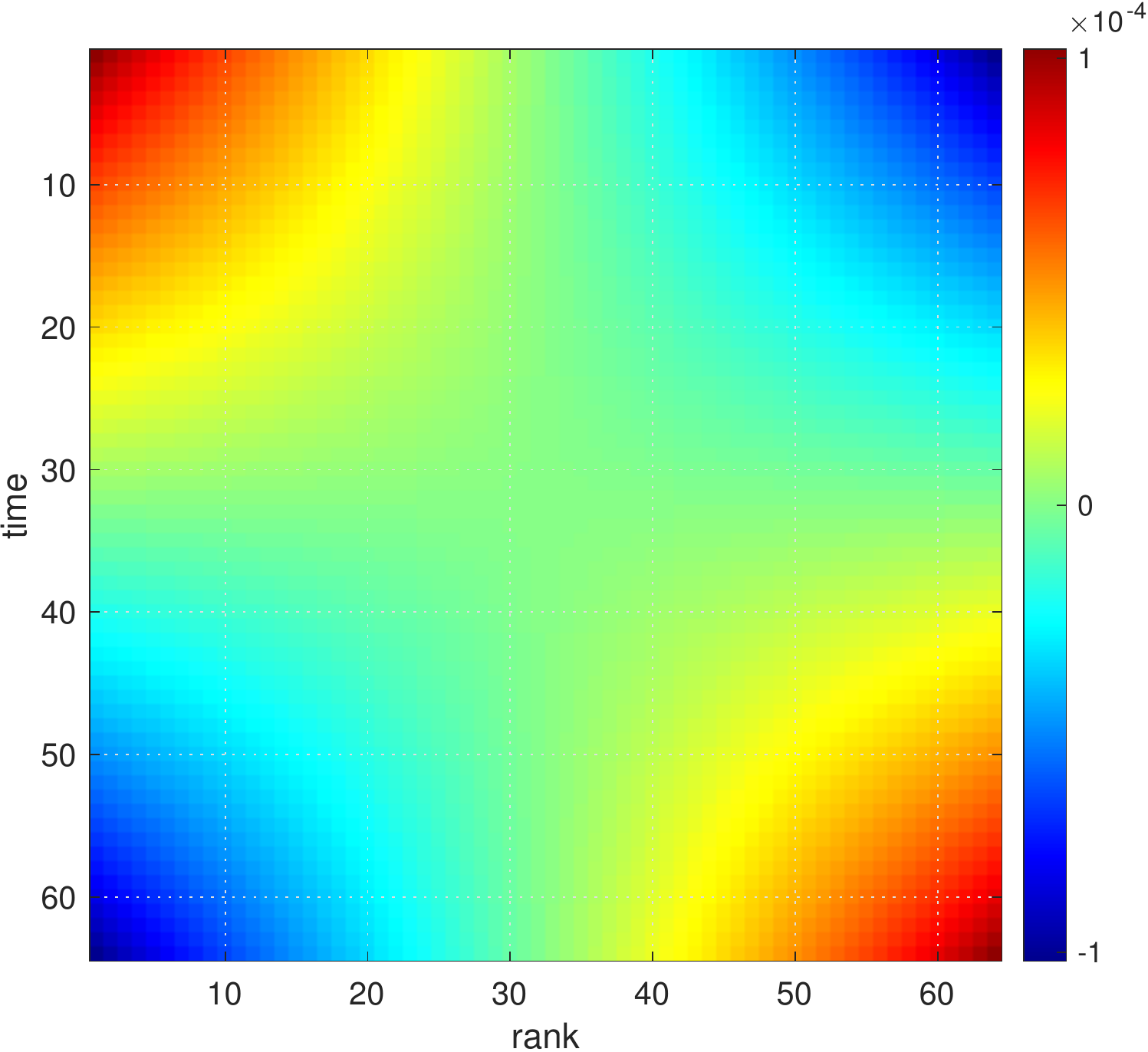}}
\caption{{\bf $\langle Q\rangle$ as a filtered product of $P$:} Here the contracting row vector ${\bf m}^{\rm T} = {\bf 1}^{\rm T} \, M$ is reshaped as a matrix to clarify its role in extracting a trend from $P$.}\label{fig:filter}
\end{figure}

\subsection{Analysis for asymptotics of $\langle Q \rangle$\label{Qasympt}}

While a derivation of (\ref{eq:meanq_asy}) is challenging, one can approach it with a simplified model developed from a computationally efficient observation about (\ref{eq:meanq}). Noting the earlier recasting of ${\bf q} = M \, {\bf p} $, if one is solely interested in $\langle Q \rangle$, this is obtained by left multiplying on both sides by ${\bf 1}^T$, a row vector of ones. We can pre-multiply at right, denoting the result as ${\bf m}^{\rm T} = {\bf 1}^{\rm T} \, M$. The result for $\langle Q \rangle$ then obtains in $K = n_T \times n_t$ flops and computation is dominated by $n_T\, n_t \log n_T$ flops for the sort operation needed for $P$. It is instructive to reconstitute ${\bf m}$ as a matrix, as shown in Fig. \ref{fig:filter}.\footnote{It is even more instructive to examine the eigenvectors and eigenvalues of ${\bf m}$ reconstituted as a matrix.} The sum of the point-wise (Hadamard) product of this field with the noisy data in $P$ is the precise content expressed in $\langle Q \rangle$, and so also then the meaning of setting $\langle Q \rangle = 0$.  
Recall that the entries in $P$ are correlated Poisson random variables. Specifically, to leading order any element $p_{i,j}$ has a correlation of $ - 1/n_T$ with all other elements in the $i^{\rm th}$ row and $j^{\rm th}$ column. For typical values of $n_T$, this is weak correlation, and so we consider instead a companion matrix $\tilde P$ populated by {\em uncorrelated} Poisson variables with the same parameter, $\lambda = n_t/n_T$. Half the elements in ${\bf m}$ are positive, the other half are the negatives of these. Accordingly, we partition the contraction ${\bf m}^{\rm T}\, \tilde{\bf p}$ into the corresponding contributions. We can use a normal approximation for the sum of uncorrelated Poisson variables with positive definite coefficients. The variance of the resulting normal random variable is 
\[
\frac{n_t}{n_T} \, \sum_{k=1}^{\lfloor n_T^2/2 \rfloor} \,
(m^{(+)}_k)^2\, .
\]
A second normal random variable from the sum with negative coefficients has exactly the same variance. Consequently, the variance of the final sum of these two is twice the above. (The means of the two are equal and opposite and so the mean of their sum is zero.) The standard deviation then follows directly. The elements $m_k^{(\pm)}$ could be expressed exactly by reference to (1) but the algebra would be formidable, to say nothing of the sum. But one can anyway observe that $m_k^{(\pm)}$ depends solely upon $n_T$ save for the overall prefactor of $1/n_t$. Here the asymptotic result that follows is 
\be\label{eq:meanq_asytilde}
\sigma_{\langle \tilde Q \rangle} \sim \frac{ 0.7015}{{n_t}^{1/2}}\,
\left [  \frac{1}{{n_T}^{1/2}}  
+ \frac{2.0313}{n_T^{3/2}} + {\cal O}( \frac{1}{{n_T}^{5/2} })\right ]\, .
\ee
With less than a two percent change in the leading order coefficient, this result is very close to (\ref{eq:meanq_asy}). The main distinction is the absence of a term of order $1/n_T$. Such a term cannot arise from the algebra that generates $m_k^{(\pm)}$. Rather it stems from the weak correlation of order $-1/n_T$ for the full problem. 

%\bibliography{IerleyKostinski}


\end{document}